# Review on quasi-2D square planar nickelates

Junjie Zhang* and Xutang Tao*

Institute of Crystal Materials, State Key Laboratory of Crystal Materials, Shandong University, Jinan, Shandong 250100, China

Email: junjie@sdu.edu.cn (J. Z.) and txt@sdu.edu.cn (X. T.)

**Abstract:** In strongly correlated materials, lattice, charge, spin and orbital degrees of freedom interact with each other, leading to emergent physical properties such as superconductivity, colossal magnetic resistance and metal-insulator transition. Quasi-2D square planar nickelates, $R_{n+1}Ni_nO_{2n+2}$ (R=rare earth, n=2, 3… ∞), are of significant interest and long sought for cuprate analogue due to the $3d^9$ electronic configuration of $Ni^+$, the same as the active ion $Cu^{2+}$ in the high-$T_c$ superconducting cuprates. The field has attracted intense attention since 2019 due to the discovery of superconductivity in thin films of $Nd_{0.8}Sr_{0.2}NiO_2$, although no superconductivity has been reported in bulk polycrystalline powders. Herein, we review the synthesis of polycrystalline powders of quasi-2D square planar nickelates through topotactic reduction of parent compounds that are synthesized via solid state reaction, precursor method, high pressure floating zone method and high-pressure flux method. We emphasize single crystal preparation using the high-pressure floating zone techniques. We discuss their crystal structure and physical properties including resistivity, magnetic susceptibility and heat capacity. We highlight the cuprate-like physics, including charge/spin stripes and large orbital polarization, identified in single crystals of $R_4Ni_3O_8$ (R=La and Pr) combining synchrotron X-ray/neutron single crystal diffraction and density functional theory calculations. Furthermore, the challenges and possible research directions of this fast-moving field in the future are briefly discussed.

## 1. Introduction

Superconductors, a class of solid materials exhibiting zero electrical resistance and total expulsion of magnetic field below a critical temperature ($T_c$), are of significant technological importance in potential applications such as zero-loss power grid, energy storage, generation of strong magnetic field for magnetic resonance imaging (MRI) and accelerators, and magnetic levitation[1]. Over one hundred years have passed since the discovery of superconductivity in the element Mercury by Heike Kamerlingh Onnes[2], who was awarded the Nobel Prize in Physics in 1913. $T_c$ has been increased from 4.2 K in Mercury[2] to 133 K in Hg–Ba–Ca–Cu–O system[3] at ambient pressure to 287.7 K (~ 15 ºC) in C-S-H system at 267 GPa[4]. However, superconductivity disappears in the hydrogen-containing materials[4-7] once high pressure is released, making them limited use in real applications. To achieve room temperature superconductivity at ambient pressure, it is crucial to understand the origin of superconductivity in cuprates and iron-based superconductors, two classes of high-$T_c$ unconventional superconductors. Unfortunately, a quantitative understanding of the nature of the superconducting state in these materials is still lacking[8, 9], hindering efficient prediction and experimental discovery of new superconducting materials.

One strategy for design and discovery of new superconductors is to search for materials that exhibit key ingredients found in the high-$T_c$ superconducting cuprates, including quasi-2D square lattice, spin one half, strong antiferromagnetic correlations, large orbital polarization of the unoccupied $e_g$ states, strong p-d hybridization, and a plethora of broken symmetry phases proximate to the parent insulating phase[10-13]. In the past years, much efforts have been devoted to low-valence square-planar nickelates[13-18], $LaNiO_3$-based heterostructures[19-22], iridate $Sr_2IrO_4$[10, 11, 23-25], and silver compounds[12, 26] as these compounds contain some or all of the electronic and structural features reported in cuprates.

Nickel oxides (nickelates) are long sought for superconductivity as a potential cuprate analogue. Nickel and copper are neighbors in the periodic table, and $Ni^{+}$ has 9 electrons in the 3*d* orbitals, which is the same as that of $Cu^{2+}$, the active ions in the high-$T_c$ superconducting cuprates[27]. Layered nickelates containing $Ni^{+}$, mainly $R_{n+1}Ni_nO_{2n+2}$, are likely to exhibit similar electronic structure as cuprates and hence superconduct. Milestones for the research of layered nickelates $R_{n+1}Ni_nO_{2n+2}$ and their related parent compounds $R_{n+1}Ni_nO_{3n+1}$[15, 28-33] are shown in **Fig. 1**. Infinite-layer $LaNiO_2$ was first synthesized by Crespin et al. in 1983 via topotactic reaction, a chemical reaction that results in a material with a crystalline orientation related to the starting product, and its structure was determined by Rietveld refinement on polycrystalline powders[28, 34]. The existence of this compound was confirmed later by Hayward et al., who reduced $LaNiO_3$ using sodium hydride[35]. Theoretically, Anisimov et al. found that $Ni^{+}$ in a square planar environment of oxygen atoms can form an $S$=1/2 antiferromagnetic insulator that may become superconducting if doped with $S$=0 $Ni^{2+}$ holes[18], pointing out the direction to look for nickelate superconductors. In contrast, Lee and Pickett highlighted the difference between infinite-layer $LaNiO_2$ and cuprates: $Ni^{1+}$ is not $Cu^{2+}$.[36] As 1/3 self-hole doped materials, the trilayer nickelates $R_4Ni_3O_8$ (R=La, Pr, Nd) were first reported by Lacorre in 1992[37]. Eighteen years later, Poltavets *et al.* reported the detailed physical properties of $La_4Ni_3O_8$ on polycrystalline powder samples, and identified a 105 K phase transition with a pronounced electronic and magnetic response[16]. Theoretical models, including an antiferromagnetic spin density wave transition[16] and a high temperature low spin state to low temperature high spin state transition[38, 39], were proposed to explain the origin of the 105 K transition. $^{139}La$ nuclear magnetic resonance measurements revealed strongly two-dimensional spin fluctuations developing into long-range order in support of the antiferromagnetic transition by Poltavets *et al.*[40] The lack of magnetic superlattice reflections in the neutron powder diffraction pattern below the 105 K transition, however, argue against an antiferromagnetic transition[16]. In contrast, a clear jump of the lattice parameter *a* and a drop of the *c* axis on cooling through the transition are consistent with the scenario of the spin state transition[41]. Besides the debate on the nature of phase transition in $La_4Ni_3O_8$, other fundamental open questions related to these nickelates include the magnetic ground state, orbital polarization, spin state, and finally superconductivity.

Single crystals of $RNiO_3$ (R=La and Pr), $R_4Ni_3O_{10}$ (R=La and Pr) and $R_4Ni_3O_8$ (R=La and Pr) were first reported by Zhang et al. from Mitchell group at Argonne National Laboratory[13, 15, 31, 32, 42-47], followed by Guo et al. from Komarek group at Max-Planck-Institute for Chemical Physics of Solids ($LaNiO_3$ single crystals[48]), Jakub et al. from Pomjakushina group at Paul Scherrer Institute ($Pr_4Ni_3O_{10}$ single crystals[49]), and Dey et al. from Klingeler group at Heidelberg University ($LaNiO_3$ single crystals[50]) using the high-pressure floating zone techniques. Additionally, synchrotron X-ray and neutron diffraction were performed on single crystals of $R_4Ni_3O_{10}$ and $R_4Ni_3O_8$ (R=La and Pr) by the Argonne group, revealing diffuse scattering peaks/streaks below their phase transition temperatures[15, 33, 43]. Collaborating with Norman theoretical team, Zhang et al. discovered that intertwined charge and spin density waves are the underlying mechanism for the metal-to-metal transition in $R_4Ni_3O_{10}$ (R=La and Pr)[43], and charge/spin stripes are the origin of the semiconductor-insulator transition in $La_4Ni_3O_8$[15, 33]. Furthermore, large orbital polarization and low spin state were revealed by Zhang et al. using synchrotron X-ray absorption spectroscopy[13]. Most importantly, the substitution of La in $La_4Ni_3O_8$ with Pr leads to metallic $Pr_4Ni_3O_8$ with large orbital polarization and suppression of charge/spin stripes, making $Pr_4Ni_3O_8$ an analogue to overdoped cuprate at the 1/3-hole doping level[13].

Motivated by the key features found in square-planar trilayer nickelate single crystals and the insulator-to-metal transition from $La_4Ni_3O_8$ to $Pr_4Ni_3O_8$ (volume contraction), Li et al. from Stanford University discovered superconductivity in Sr-doped $NdNiO_2$ thin films with $T_c \sim$ 9-15 K[51], boosting intense theoretical investigations[52-68]. Later, a superconducting dome of Sr-doped $NdNiO_2$ was reported by Li et al.[69] and Zeng et al.[70] More recently, $Pr_{1-x}Sr_xNiO_2$ were found to superconduct with a superconducting dome ranging from 0.12 to 0.28[71, 72].

In this highlight, we review the recent advance in the low-valence square-planar nickelates, a fast-moving field following the discovery of superconductivity in thin films. The outline of this review is as follows: First, we discuss the polycrystalline synthesis. Second, we introduce the crystal structure of the low-valence nickelates as a function of Ni-O layer. Next, we summarize single crystal growth of these materials using flux method and high-pressure floating zone techniques. We then discuss the physical properties, followed by the cuprate-like physics such as charge and spin stripes, large orbital polarization, and low spin state. Finally, we conclude and outlook this fast-moving field.

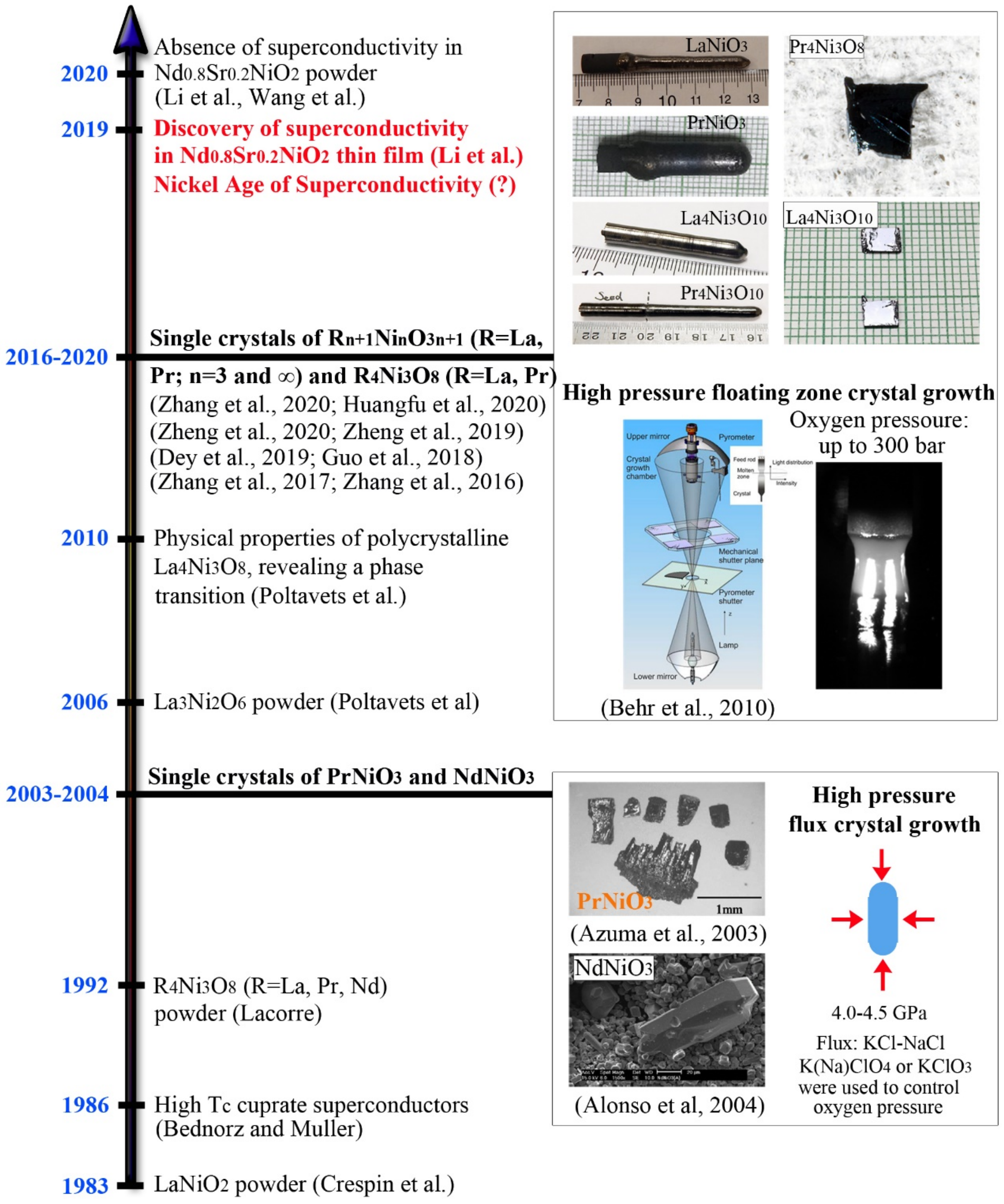

**Fig. 1** Milestones for the research of layered nickelates $R_{n+1}Ni_nO_{2n+2}$ and their parent compounds $R_{n+1}Ni_nO_{3n+1}$. (Li et al., 2020[73]; Wang et al., 2020[74]; Li et al., 2019[51]; Zhang et al., 2020[32]; Huangfu et al., 2020[49]; Zheng et al., 2020[42]; Zheng et al., 2019[44]; Dey et al., 2019[50]; Guo et al., 2018[48]; Zhang et al., 2017[13, 31]; Zhang et al., 2016[15]; Poltavets et al., 2010[16]; Behr et al., 2010[75]; Poltavets et al., 2006[76]; Alonso et al., 2004[29]; Azuma et al., 2003[77]; Lacorre, 1992[37]; Bednorz and Muller, 1986[78]; Crespin et al., 1983[28])

## 2. Polycrystalline synthesis

Low-valence square planar nickelates include $R_{n+1}Ni_nO_{2n+2}$ (R=La, Pr and Nd, n=2, 3 …), where Ni has a valence of nominal $(1+1/n)+$. These oxides cannot be directly synthesized. Instead, a two-step method was utilized[13, 15, 28, 35, 37, 51, 76, 79-81]. Specifically, parent compounds $R_{n+1}Ni_nO_{3n+1}$, which belong to the series of Ruddlesden-Popper (R-P) phases, are synthesized first, and then these R-P phases are topotactic reduced using hydrogen gas or metal hydrides such as NaH and $CaH_2$ (**Fig. 2**). **Table I** summaries the synthesis conditions of the square planar $R_{n+1}Ni_nO_{2n+2}$ (n=2, 3, ∞) and corresponding parent compound.

**Synthesis of R-P $R_{n+1}Ni_nO_{3n+1}$ (n=2, 3, ∞).** There are multiple ways to synthesize the parent compounds: (i) Solid state reaction. (ii) Precursor method. (iii) Flux method. (iv) High pressure floating zone technique. For n=∞, Method (i-iv) has been reported. For synthesis of bulk $Nd_{1-x}Sr_xNiO_3$ samples, all $x$ = 0.1, 0.2 and 0.4 samples were found to contain NiO impurities[73, 74]. In contrast, Method (i, iii, iv) was used to synthesize n=3 compounds, and Method (i, iii) were reported to prepare $La_3Ni_2O_7$, whereas $Pr_3Ni_2O_7$ and $Nd_3Ni_2O_7$ have not been reported.

**Topotactic reduction**. To reduce the parent compounds, hydrogen gas or metal hydrides were utilized. All of $R_{n+1}Ni_nO_{3n+1}$ (n=2, 3, ∞) can be reduced using hydrides, but only $LaNiO_2$ and $R_4Ni_3O_8$ (R=La, Pr and Nd) were successfully reduced using hydrogen or diluted hydrogen gas.

(i) Hydrogen gas. $LaNiO_2$ and $R_4Ni_3O_8$ (R=La, Pr and Nd) were reduced from corresponding parent compounds by hydrogen gas[13, 15, 28, 34, 37, 80]. The synthesis of $LaNiO_2$ was reported by Crespin et al. in 1983, in which $LaNiO_2$ was obtained using a controlled hydrogen reduction loop at moderate temperature (570-670 K)[28, 34]. However, the existence of $LaNiO_2$ was questioned due to several unsuccessful attempts to reproduce Crespin et al.'s result[35, 80, 82]. In 1992, the square planar trilayer nickelates $R_4Ni_3O_8$ were prepared by Lacorre using hydrogen gas[37]. Pure or diluted hydrogen can be used to reduce the $R_4Ni_3O_{10}$ parent compounds to $R_4Ni_3O_8$ in the form of polycrystalline powders and single crystals[15].

(ii) Metal hydrides. Hayward et al. synthesized infinite-layer $LaNiO_2$ using solid sodium hydride in a sealed evacuated tube at 460-480 K[35], confirming Crespin's finding[28]. In this method, polycrystalline $LaNiO_3$ was thoroughly ground with NaH, sealed in an ampule, and then heated. After reaction, the products were washed with $CH_3OH$ to remove the reaction byproduct NaOH and any unreacted NaH[35]. Similar method was applied to prepare $NdNiO_2$ by Hayward et al.[35, 81], the trilayer $La_4Ni_3O_8$ by Blakely et al.[83] and $Nd_4Ni_3O_8$ by Li et al.[84], and the bilayer $La_3Ni_2O_6$ by Poltavets et al.[76] Recently, thin films of $RNiO_2$ and 20% Sr doped $RNiO_2$ (R=Pr, Nd) were prepared using hydrides, and the doped films were reported to superconduct below 15 K[51, 69, 70, 72]. Recently, polycrystalline samples of $RNiO_2$ and 20% Sr doped $RNiO_2$ (R=Pr, Nd) have been synthesized; however, superconductivity was not observed[73, 74]. Notably, all of Sr doped $NdNiO_2$ powders contain some amounts of nickel impurity[73, 74]. The underlying mechanism of reduction may be solid state reaction or via hydrogen gas, and the successful reduction from $Nd_{1-x}Sr_xNiO_3$ to $Nd_{1-x}Sr_xNiO_2$ using physically separated $CaH_2$ supports the latter[74].

**Table I**. Synthesis of bulk samples of $R_{n+1}Ni_nO_{3n+1}$ and $R_{n+1}Ni_nO_{2n+2}$ (n=2, 3, ∞)

| **Materials** | **Nickel valence** | **Space group** | **Reduction from parent compounds** | | **Parent compounds** | **Nickel valence** | **Space group** | **Preparation method of parent compounds** | | | |
|---|---|---|---|---|---|---|---|---|---|---|---|
| | | | **$H_2$ gas** | **Metal hydrides** | | | | **Solid state reaction** | **Flux method** | **Precursor method** | **Floating zone technique** |
| $LaNiO_2$ | 1+ | *P4/mmm* | $H_2$, 310 °C, 25 h (Ref. [28, 34, 80]) | NaH, 190-210 °C, 3×3 days (Ref.[35]) | $LaNiO_3$ | 3+ | $R\bar{3}c$ | 60k bar $O_2$ at 950 °C (Ref.[85]) | 800-850 °C in $Na_2CO_3$ flux under 0.2 bar $O_2$, (Ref.[86, 87]) | Nitrate route (Ref.[88, 89]), Sol-gel and calcine in air (Ref.[90]) | 30-150 bar $O_2$, 3-7 mm/h, 10-20 rpm (Ref.[31, 32, 46, 48, 50]) |
| $PrNiO_2$ | 1+ | *P4/mmm* | × | × | $PrNiO_3$ | 3+ | *Pbnm* | 60k bar $O_2$ at 950 °C (Ref.[85]) | 1450-1250 °C in KCl, NaCl, $KClO_4$, $NaClO_4$@4.5 GPa (Ref.[77]) | Nitrate route (Ref.[88, 89]) | 295 bar $O_2$, 5 mm/h, 15 rpm (Ref.[44]) |
| $NdNiO_2$ | 1+ | *P4/mmm* | × | NaH, 160-200 °C, 4×3 days (Ref.[35, 81]) | $NdNiO_3$ | 3+ | *Pbnm* | 60k bar $O_2$ at 950 °C (Ref.[85]), OR $Nd_2NiO_4$, NiO and $KClO_4$ at 1000 °C@2 GPa (Ref. [73]) | 900-600 °C in $KClO_3$ and KCl@ 5 GPa (Ref.[29]) | Nitrate route (Ref.[89, 91]) or sol-gel (Ref.[91]), calcine in 1 bar of $O_2$ | × |
| $Nd_{0.9}Sr_{0.1}NiO_2$ | 1.1+ | *P4/mmm* | × | $CaH_2$, 280-285 °C, 20-48 h (Ref. [73, 74]) | $Nd_{0.9}Sr_{0.1}NiO_3$ | 3.1+ | *Pbnm* | × | × | Citrate-nitrate auto-combustion synthesis with pellets fired multiple times at 1000 °C under 150–160 bar (Ref. [74]) | × |
| $Nd_{0.8}Sr_{0.2}NiO_2$ | 1.2+ | *P4/mmm* | × | | $Nd_{0.8}Sr_{0.2}NiO_3$ | 3.2+ | *Pbnm* | Mixture of $R_{2-2x}Sr_{2x}NiO_4$ (R=Nd, Sm, *x*=0.2, 0.4), NiO and $KClO_4$ at 1000 °C @ 2 GPa (Ref. [73, 92]) | × | | × |
| $Nd_{0.6}Sr_{0.4}NiO_2$ | 1.4+ | *P4/mmm* | × | | $Nd_{0.6}Sr_{0.4}NiO_3$ | 3.4+ | *Pbnm* | | × | × | × |
| $Sm_{0.8}Sr_{0.2}NiO_2$ | 1.2+ | *P4/mmm* | × | $CaH_2$, 340 °C, 10 h (Ref. [92]) | $Sm_{0.8}Sr_{0.2}NiO_3$ | 3.2+ | *Pbnm* | | × | × | × |
| $La_4Ni_3O_8$ | 1.33+ | *I4/mmm* | $H_2$ or 6% $H_2/N_2$, 1-3 days, 300 °C (Ref. [37]), 4%$H_2$/Ar, 350 °C, 5 days (Ref.[15]) | NaH, 200 °C, 240 h (Ref. [83]) | $La_4Ni_3O_{10}$ | 2.67+ | $P2_1/a$, *Bmab* | 0.2 bar of $O_2$ at 1080 °C (Ref.[93]) | × | Nitrate route and sintering in air (Ref.[93]), Sol-gel and sinter in air (Ref.[93]) | 20 bar $O_2$, 4-6 mm/h, 20 rpm (Ref.[15, 32, 33]) |
| $Pr_4Ni_3O_8$ | 1.33+ | *I4/mmm* | | - | $Pr_4Ni_3O_{10}$ | 2.67+ | $P2_1/a$ | × | × | | 140 bar $O_2$, 4-6 mm/h, 20 rpm (Ref.[13, 32, 49]) |
| $Nd_4Ni_3O_8$ | 1.33+ | *I4/mmm* | $H_2$ or 6% $H_2/N_2$, 1-3 days, 300 °C (Ref. [37]), $H_2$, 320 °C (Ref. [94]) | $CaH_2$, 280 °C, 20 h (Ref. [84]) | $Nd_4Ni_3O_{10}$ | 2.67+ | $P2_1/a$ | × | × | | × |
| $La_3Ni_2O_6$ | 1.5+ | *I4/mmm* | × | $CaH_2$, 350 °C, 4 days (Ref.[76]) | $La_3Ni_2O_7$ | 2.5+ | *Cmca* | 0.2 bar of $O_2$ at 1150 °C (Ref.[93]) | × | | × |

### 3. Crystal structure

Low-valence square planar nickelates, $R_{n+1}Ni_nO_{2n+2}$ (R=La, Pr and Nd, n=2, 3 …), possess quasi-two-dimensional crystal structure with nickel atoms coordinated by four oxygen atoms, which is isostructural to the electron doped cuprates such as $R_{2-x}Ce_xCuO_4$ (R=Nd, Pr, Sm and Eu)[95].

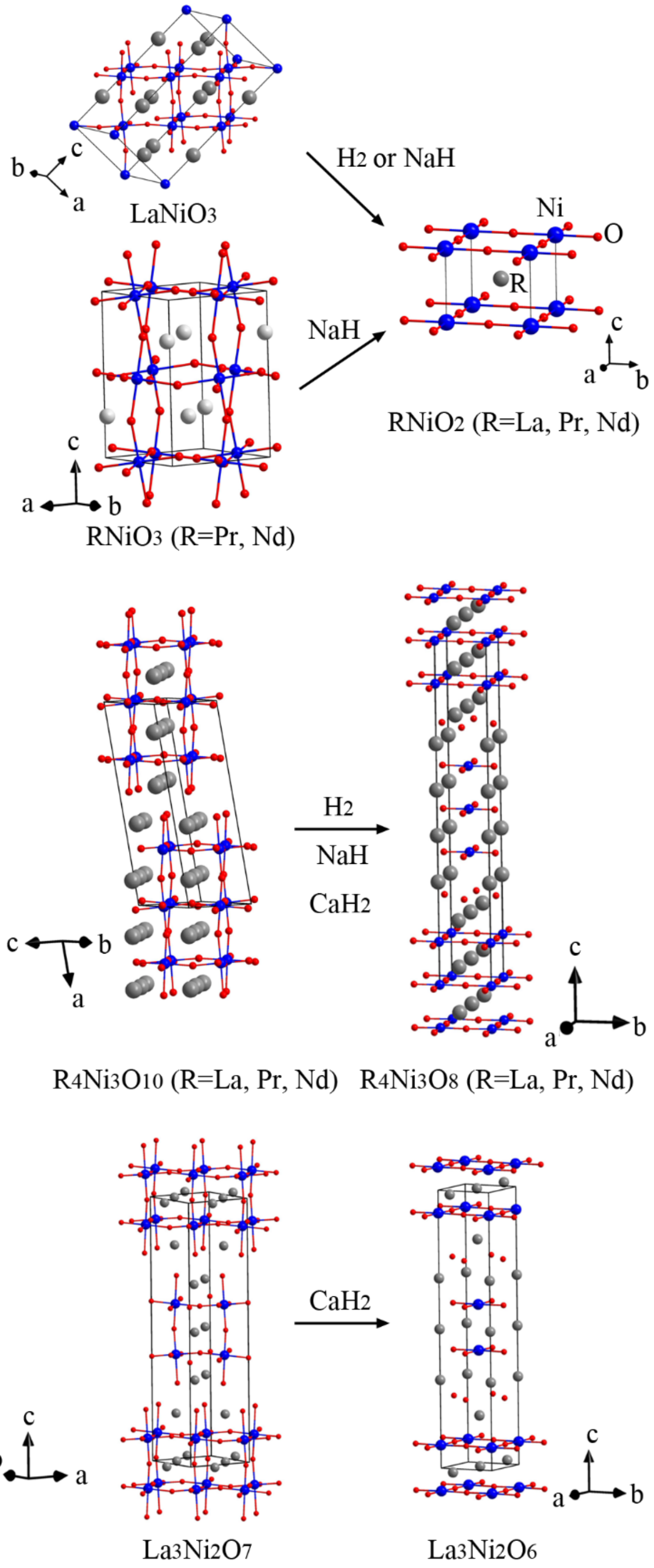

**Fig. 2** Topotactic reaction converting the Ruddlesden-Popper $R_{n+1}Ni_nO_{3n+1}$ to the quasi-2D square planar nickelates $R_{n+1}Ni_nO_{2n+2}$[13, 15, 31, 32, 76, 80, 96].

The crystal structure of $LaNiO_2$ and $NdNiO_2$, as shown in **Fig. 2**, were obtained by Rietveld refinements on neutron powder diffraction data by Crespin et al.[34, 80] and Hayward et al.[35, 81], respectively. Both $LaNiO_2$ and $NdNiO_2$ crystallize in the *P4/mmm* space group with lattice parameters of *a*=3.959(1) Å and *b*=3.375(1) Å for $LaNiO_2$[80], and *a*=3.9208(3) Å and *b*=3.2810(8) Å for $NdNiO_2$[81]. There is one rare-earth (La or Nd) atom (Wyckoff position 1*d*), one Ni atom (1*a*), and one oxygen atom (2*f*) in the asymmetric unit, each of which have an occupancy of 1. The bond lengths for Ni-O are 1.980(1) Å for $LaNiO_2$[80] and 1.9604(2) Å for $NdNiO_2$[81]. Nickel atoms are four coordinated by oxygen atoms, forming square planar geometry. These square planar form infinite sheets and stacking along the *c* axis. It is worth noting that Hayward et al. refined their diffraction data using mixed phases consisting of perfect infinite-layer phase $RNiO_2$ (R=La and Nd), defective regions which have interlamellar oxide ion defects ($RNiO_{2.09}$), and Ni metal. Such an observation was not found in hydrogen reduced samples by Crespin et al.[80] or $CaH_2$ reduced samples by Wang et al.[74]

The crystal structure of $R_4Ni_3O_8$ were reported by Lacorre et al,[37, 94] Poltavets et al[16, 79], Li et al,[84] and Huangfu et al[97] from Rietveld refinement on powder diffraction data, and by Zhang et al[13, 15] on single crystal X-ray diffraction data. They crystalize in the *I4/mmm* space group with *a*=3.9700(5) Å, *c*=26.092(3) Å at 296 K for $La_4Ni_3O_8$[15], *a*=3.9347(1) Å, *b*=25.4850(9) Å at 300 K for $Pr_4Ni_3O_8$[13], and *a*=3.9142(9) Å, *c*=25.296(7) Å for $Nd_4Ni_3O_8$[84]. Two R atoms (Wyckoff position 4*e* and 4*e*), two Ni atoms (2*a* and 4*e*) and three O atoms (4*c*, 8*g* and 4*d*) were found in the asymmetric unit with site occupancy of 1 for all of them. The structure of $R_4Ni_3O_8$ (R=La and Pr) can be viewed as fluorite blocks sandwiched by trilayer blocks which consist of three infinite layers stacking on top of each other[13, 15], as shown in **Fig. 2**. The rare-earth atoms are coordinated by eight oxygen atoms, and nickel atoms are surrounded by four oxygen atoms with Ni-O bond lengths in the range of 1.9850(2)-1.9852(3) Å for $La_4Ni_3O_8$[15] and 1.96735(5)-1.96760(11) Å for $Pr_4Ni_3O_8$[13].

The crystallographic data of the bilayer $La_3Ni_2O_6$ were obtained from Rietveld refinements on polycrystalline powder neutron diffraction data at 300 K[76]. It belongs to *I*4/*mmm* space group with *a*=3.9686(1) Å and *c*=19.3154(6) Å. There are two La atoms (Wyckoff position 2*b* and 4*e*), one Ni (4*e*), and two O (4*d* and 8*g*) in the asymmetric unit. The structure can be described as intergrowth of double infinite layer blocks ($La/NiO_2/La/NiO_2$) and fluorite blocks ($La/O_2/La$) (see **Fig. 2**). The Pr and Nd analogues have not been reported yet.

## 4. Single crystal preparation

Single crystals of low valent nickelates $R_{n+1}Ni_nO_{2n+2}$ are obtained through reduction of corresponding high valent parent compound $R_{n+1}Ni_nO_{3n+1}$ (R=rare earth; n=2, 3 …∞). Crystal growth of high valent nickelates has been very challenging. Up to date, high pressure flux method[29, 77] and high oxygen pressure floating zone techniques[13, 15, 31, 32, 42, 44, 46, 48-50] have been utilized to grow single crystals.

Bulk $RNiO_2$ (R=La, Pr, Nd) single crystals have not been reported, although single crystals of $PrNiO_3$[77], $NdNiO_3$[29] and $LaNiO_3$[31, 42, 46, 48, 50] were grown (see **Fig. 1**). Glossy and dark brown single crystals of $PrNiO_3$ with typical dimensions of 0.5×0.5×0.5 $mm^3$ were grown by slowly cooling the solution of $Pr_6O_{11}$, NiO, KCl, NaCl, $KClO_4$, and $NaClO_4$ at 4.5 GPa, where KCl and NaCl worked as flux, and $KClO_4$ and $NaClO_4$ served as oxidizers to control oxygen pressure[30, 77]. Similarly, needle shaped single crystals of $NdNiO_3$ with average size of 50-100 $\mu$m were grown from solution of $Nd(OH)_3$, $Ni(OH)_2$ and $KClO_3$ under a pressure of 4 GPa[29]. Like $PrNiO_3$, $KClO_3$ served as an oxidizer to control the oxygen pressure. The preparation of other $R_{n+1}Ni_nO_{3n+1}$ (R=La, Pr, and Nd, n=2, 3, ∞) bulk single crystals has not been reported using flux method.

An alternative method is the high oxygen pressure floating zone technique (see **Fig. 1**)[31, 42, 44, 46, 48, 50]. The floating zone crystal growth technique has played important roles in both technological advances and the development of pristine crystals[98]. The advantages of this technique include (1) the ability to create a molten zone via surface tension so that a container/crucible is not needed, which avoids contamination from container/crucible where impurities are often introduced, and (2) the ability to remove impurities by zone refining, so that ultrahigh purity sample can be prepared. By applying high pressure of gases (e.g. oxygen, nitrogen, etc.), floating zone techniques have brought new opportunities in exploration of novel quantum states and become an expanding frontier in new materials synthesis and discovery[13, 15, 33, 47, 99-104]. Specifically, most inert (e.g. Ar) and reactive gases (e.g. $N_2$, $O_2$) transition to liquids or supercritical fluids, where liquid and gas cannot be distinguished, in the range of 10-10000 bar. For example, oxygen gas become supercritical fluid at around 50 bar[105] (see **Fig. 3a**). For a reaction between a solid/liquid sample material and a gas/fluid atmosphere, changes in pressure up to 10000 bar result in significant changes in the properties of the fluid that impacts the chemistry of the sample[105].

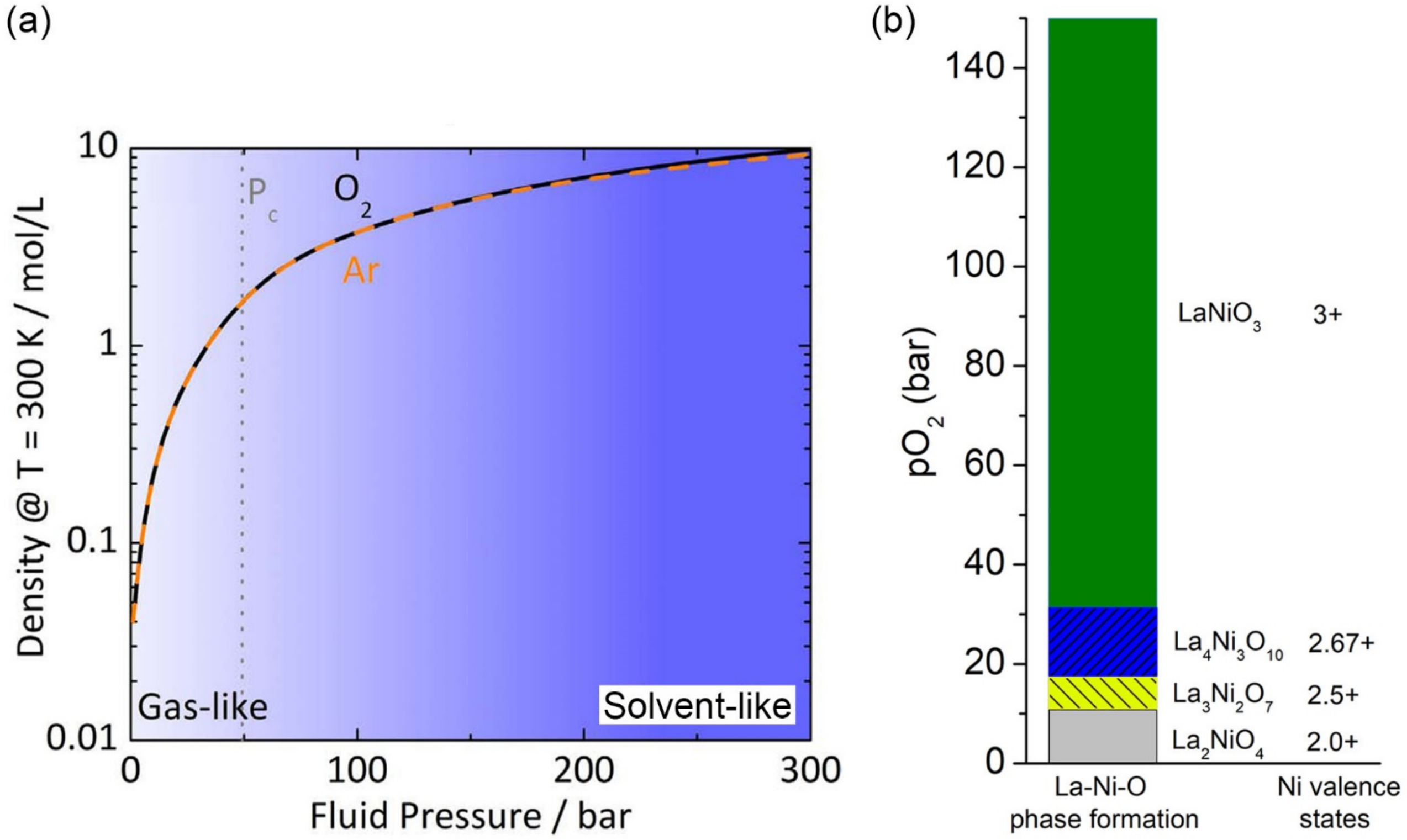

**Fig. 3** (a) Gas-like and solvent-like state in oxygen as a function of pressure[105] and (b) Schematic drawing of empirical phase predominance for the La-Ni-O system as a function of $pO_2$[32].

Bulk crystal growth of the correlated metal $LaNiO_{3-\delta}$ (see **Fig. 1**) was chronologically reported by Zhang et al.[31, 46], Guo et al.[48], Dey et al.[50] , and Zheng et al.[42] using the high-pressure floating zone furnaces supplied by SciDre GmbH, Dresden. In 2017, Zhang et al. first grew $LaNiO_{2.985}$ under an oxygen pressure of 30-50 bar with a traveling rate of 3 mm/h[31]. Later, Guo et al. reported the growth of $LaNiO_3$ under oxygen pressure of 130-150 bar with a growth speed of 6-7 mm/h[48]. The crystals exhibit an anomaly at 157 K in both magnetic susceptibility and heat capacity, which was missing in Zhang et al.'s samples[31]. To check the anomaly, Zhang et al. grew $LaNiO_3$ crystals at oxygen pressure of 130 bar[46], and Wang et al. found that oxygen deficient $LaNiO_{2.5}$ with $T_N$~152 K exist as dilute secondary phase in the as-grown crystals, providing an alternative explanation for the anomaly observed by Guo et al.[48] Recently, cm-sized single crystals of $LaNiO_3$ were grown at oxygen pressures of 40 and 80 bar by Dey et al.[50], whose data indicate that long-range magnetic order is not intrinsic to $LaNiO_3$. More recently, Zheng et al.[42] prepared $LaNiO_3$ at 149 bar $O_2$ and found a radial gradient in the as-grown sample with $LaNiO_{2.5}$ in the center and $LaNiO_3$ in the periphery of the boule. The as-grown $LaNiO_{2.5}$ can be fully oxidized to $LaNiO_3$ under high oxygen pressure, and then reduced back to $LaNiO_{2.5}$ using $H_2$. Surprisingly, these research groups did not reduce $LaNiO_{3-x}$ to $LaNiO_2$.

With the decreasing of rare earth ionic size, the oxygen pressure needed to stabilize a perovskite phase increases. $PrNiO_{3-\delta}$ single crystals with dimensions of ~1 $mm^3$ were grown by Zheng et al.[44] at oxygen pressure of 295 bar using a 300-bar high-pressure optical image floating zone furnace (see **Fig. 1**). The formation of $RNiO_3$ (R≤Nd, $n$=3 and ∞) requires oxygen pressure of more than 300 bar, which is beyond the limit of current commercially high-pressure floating zone furnace[106].

Besides n=∞, bulk single crystals of $R_{n+1}Ni_nO_{3n+1}$ (R=La and Pr, $n$=3) have also been successfully grown using the high oxygen pressure floating zone techniques. For La-Ni-O system, a schematic drawing of empirical phase predominance as a function of oxygen pressure with starting materials of $La_2O_3$:NiO=2:3 is shown in **Fig. 3b**[32]. The oxygen pressure for stabilizing a phase increases with *n*, consistent with the increase of nickel oxidization state from $Ni^{2+}$ in $La_2NiO_4$ to $Ni^{2.5+}$ in $La_3Ni_2O_7$ to $Ni^{2.67+}$ in $La_4Ni_3O_{10}$ finally to $Ni^{3+}$ in $LaNiO_3$.

Single crystals of $R_4Ni_3O_{10}$ (R=La, Pr, see **Fig. 1**)) were grown under oxygen pressure of 20 bar for $La_4Ni_3O_{10}$ and 140 bar for $Pr_4Ni_3O_{10}$ using the a 150-bar high-pressure floating zone furnace by Zhang et al.[32] $La_4Ni_3O_{10}$ crystals were grown directly from the sintered rod. For $Pr_4Ni_3O_{10}$, a two-step method was utilized. The first step was to densify the rod using 30-50 mm/h under an oxygen pressure of 140 bar, and the second step was to grow single crystals at the same pressure with slow travelling rates. Postgrowth cooling rate is an important parameter in obtaining the thermodynamically stable phase and trapping metastable phases. It has been found that $R_4Ni_3O_{10}$ (R=La and Pr) crystallizes in the monoclinic $P2_1/a$ ($Z$ = 2) space group at room temperature, and a metastable orthorhombic phase (*Bmab*) can be trapped by postgrowth rapid cooling. Single crystal growth of $Pr_4Ni_3O_{10}$ was also reported by Jakub et al. from Paul Scherrer Institute, who successfully grew single crystals with dimensions up to 3.3 mm×1.6 mm×0.3 mm.[49]

Recently, single crystals of $R_4Ni_3O_8$ (R=La, Pr) with typical dimensions of 1-2 $mm^2$ × 0.5 mm were reported (see **Fig. 1**) by reducing corresponding $R_4Ni_3O_{10}$ (R=La, Pr) crystals using hydrogen gas by Zhang et al.[13, 15, 33]. The topotactic reduction was carried out at 350 °C flowing 4% $H_2$/Ar gas for five days. These crystals are fragile because of strains and microcracks that develop during the reduction process. The availability of these crystals provides an ideal platform to unlock the origin of the phase transition in $La_4Ni_3O_8$, and to investigate whether the key ingredients of high temperature superconductivity in cuprates exist in the low-valence layered nickealtes.

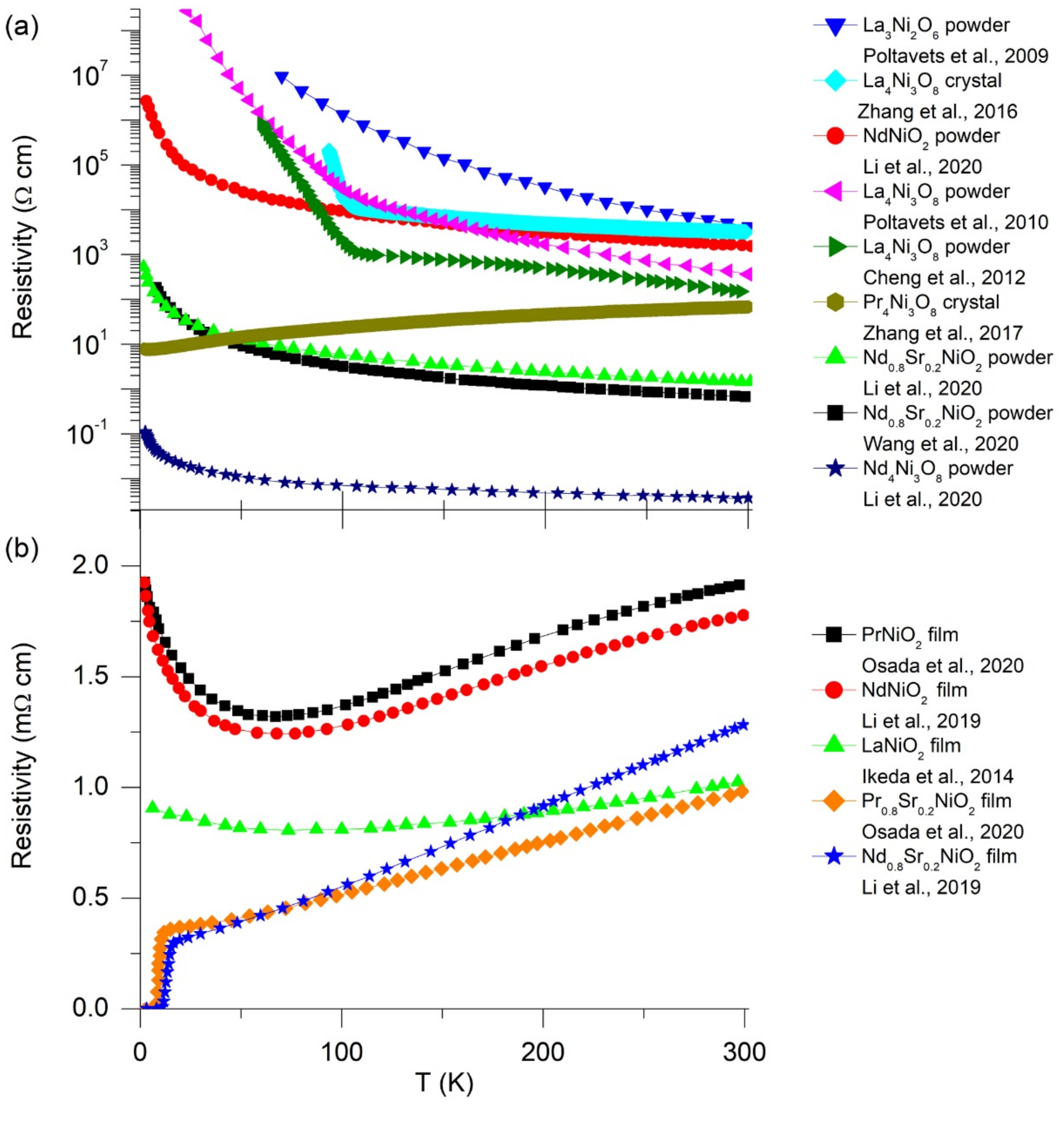

**Fig. 4** Resistivity of undoped and doped $R_{n+1}Ni_nO_{2n+2}$ (R=La, Pr and Nd, n=2, 3, ∞) in the bulk[13, 15-17, 41, 73, 74, 84] and thin-film form[51, 72, 107]. Poltavets et al., 2009[17]; Zhang et al., 2016[15]; Li et al., 2020[73]; Poltavets et al., 2010[16]; Cheng et al., 2012[41]; Zhang et al., 2017[13]; Li et al., 2020[73]; Wang et al., 2020[74]; Li et al., 2020[84]; Osada et al., 2020[72]; Li et al., 2019[51]; Ikeda et al., 2014[107]; Osada et al., 2020[72]; Li et al., 2019[51].

## 5. Physical properties of square planar nickelates

Measurements on physical properties including resistivity, magnetic susceptibility, and heat capacity were reported on polycrystalline powders, single crystals and thin films.

**Resistivity**. **Figs.4a,b** summarizes the resistivity of undoped and doped $R_{n+1}Ni_nO_{2n+2}$ (R=La, Pr and Nd, n=2, 3, ∞) in the bulk and thin film form, respectively. We first discuss the resistivity of bulk samples shown in **Fig.4a**. All of them, except $Pr_4Ni_3O_8$, are semiconducting. The resistivity of polycrystalline $LaNiO_2$ is 10(±5) Ω cm at 300 K, but no temperature dependence was reported[35]. Bulk $NdNiO_2$ samples exhibit semiconducting/insulating behavior with resistivity ranging from ~1500 Ω cm at 300 K to ~$2.69\times10^6$ Ω cm at 50 K[73]. Hole doping in $NdNiO_2$ by substituting Nd with Sr results in more metallic samples but they do not superconduct[73, 74]. There is no report on the resistivity of $PrNiO_2$ bulk samples. For *n*=3 samples, Lacorre initially had a note that $R_4Ni_3O_8$ (R=La, Pr and Nd) are black semiconductors[37]. Temperature dependent resistivity for $La_4Ni_3O_8$ was reported by Poltavets et al., who observed a semiconducting behavior and a phase transition at around 105 K[16]. Such a behavior was verified by single crystal data[15]. In sharp contrast, $Pr_4Ni_3O_8$ is metallic as the in-plane resistivity measured on single crystals decrease with decreasing temperature[13]. This is understandable, considering the volume shrinks from $La_4Ni_3O_8$ to $Pr_4Ni_3O_8$, and the compression of $La_4Ni_3O_8$ under high pressure leads to less insulating behavior[41]. Along this line, one would expect that $Nd_4Ni_3O_8$ is metallic in the *ab* plane, and this remains to be tested. Recently, measurements on polycrystalline $Nd_4Ni_3O_8$ indicates semiconducting behavior[84, 108]. It is possible that the in-plane resistivity of $Nd_4Ni_3O_8$ is metallic and out-of-plane is insulating, and the bulk resistivity is dominated by the latter. Another possibility is that $Pr_4Ni_3O_8$ behaves different from $La_4Ni_3O_8$ and $Nd_4Ni_3O_8$, like what has been reported in cuprates, *i.e.*, Pr is the only case that forms the $RBa_2Cu_3O_7$ structure but does not superconduct[109]. It is worth noting that the valence state of Pr is 3+ in $Pr_4Ni_3O_8$, as evidenced by X-ray absorption spectroscopy[13]. For *n*=2, the resistivity of $La_3Ni_2O_6$ is 4000 Ω cm at 300 K, which increases as a function of decreasing temperature, indicating semiconducting behavior[17].

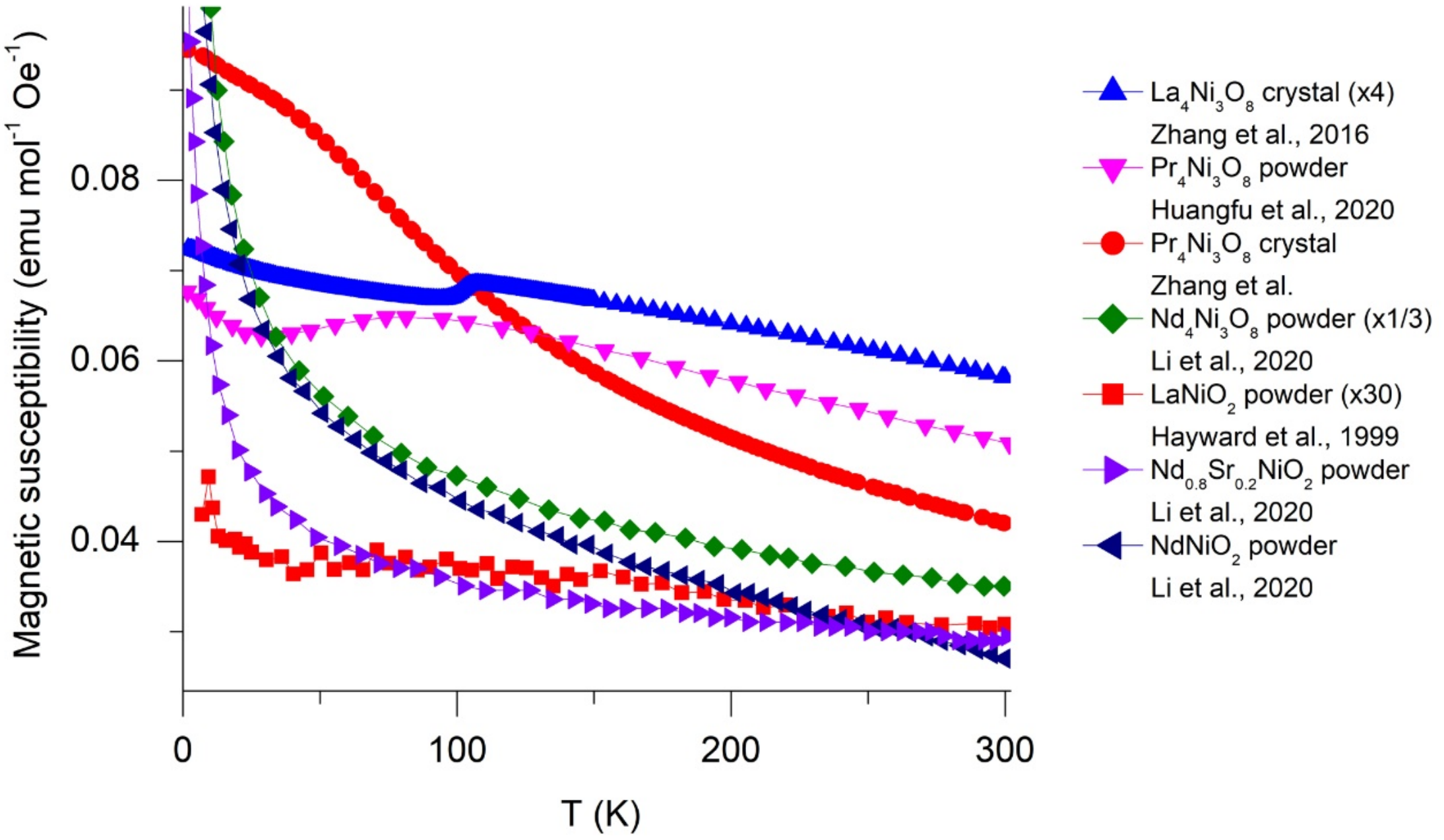

**Fig. 5** Magnetic susceptibility of undoped and doped $R_{n+1}Ni_nO_{2n+2}$ (R=La, Pr and Nd, n=2, 3, ∞). Zhang et al.,

2016[15]; Huangfu et al., 2020[97]; Zhang et al.; Li et al., 2020[84]; Hayward et al., 1999[35]; Li et al., 2020[73]; Li et al., 2020[73].

Compared with bulk materials, *c*-axis orientated thin films of $RNiO_2$ (R=La, Pr, Nd) are metallic[51, 72, 107, 110] with an upturn at around 50 K. It has also been shown that low-temperature annealing with $CaH_2$ first reduces a single-crystalline epitaxially grown $LaNiO_3$ thin film to a *c*-axis orientated $LaNiO_2$ thin film and then changes to an *a*-axis orientated film consisting of twin structures[111]. More importantly, hole doping in $RNiO_2$ (R=Pr, Nd) leads to lower resistivity, and samples with 20% hole doping are superconducting below ~15 K[51, 72].

The difference in resistivity between bulk samples and thin films is not unexpected. First, there is strong anisotropy in resistivity for such quasi-2D square planar nickelates, *i.e.*, out-of-plane resistivity is probably several orders of magnitude higher than that of in-plane resistivity (e.g. 2 orders in manganites[112] and 2-4 orders in cuprates[113]). For random orientated bulk samples, resistivity is dominated by the out-of-plane component, thus showing overall insulating behavior. Second, there is lattice mismatch between substrates and target results in strain/stress in thin films. The in-plane lattice parameter is 3.9208(3) Å for $NdNiO_2$[81], which is 0.26% larger than 3.91Å for (001) $SrTiO_3$, leading to compressive strain in the thin films. Similar compressive strain exists in $Nd_{0.8}Sr_{0.2}NiO_2$[51] thin films. Considering the ion sizes $Nd^{3+}<Pr^{3+}<La^{3+}$, the *a*-axis parameters of $PrNiO_2$[72] and $Pr_{0.8}Ni_{0.2}NiO_2$[72] are larger than that of $NdNiO_2$, resulting in larger compressive in-plane strain. Third, the reported polycrystalline samples contain Ni impurity and 5-11% Ni deficiency in the infinite layer[73, 74], which mask the intrinsic properties of these low valence infinite-layer nickelates. High quality samples are crucial for exploring the intrinsic physical properties including superconductivity.

**Magnetic susceptibility**. The intrinsic magnetic properties of $R_{n+1}Ni_nO_{2n+2}$ (R=La, Pr and Nd, n=2, 3, ∞) has been challenging [15, 16, 35, 81] considering the ferromagnetic background from Ni impurity reduced from NiO[73, 74]. Polycrystalline $LaNiO_2$ is paramagnetic with a slope change at ~150 K (see **Fig. 5**), whose origin remains an open question[35]. Polycrystalline $NdNiO_2$ is paramagnetic[73]. For *n*=3, both polycrystalline powder and single crystals of $La_4Ni_3O_8$ show an antiferromagnetic magnetic order at 105 K[16, 33, 114]. In contrast, $Pr_4Ni_3O_8$ single crystals and $Nd_4Ni_3O_8$ polycrystalline powders do not show any sign of long range magnetic order down to 2 K[13, 84, 97]. Recently, Huangfu et al. reported a spin glass behavior with a freezing temperature of 68 K in $Pr_4Ni_3O_8$ polycrystalline powders[97]. The bilayer $La_3Ni_2O_6$ polycrystalline powders show no magnetic order down to 4 K[17].

To reduce the influence of the ferromagnetic background of Ni impurity, two methods were reported: (i) magnetic susceptibility of $LaNiO_2$ and $La_4Ni_3O_8$ were extracted via linear fit to the magnetization at high fields[15, 35], and (ii) the product powder containing parent phases and NiO was washed with dilute $HNO_3$ solution before reduction[115].

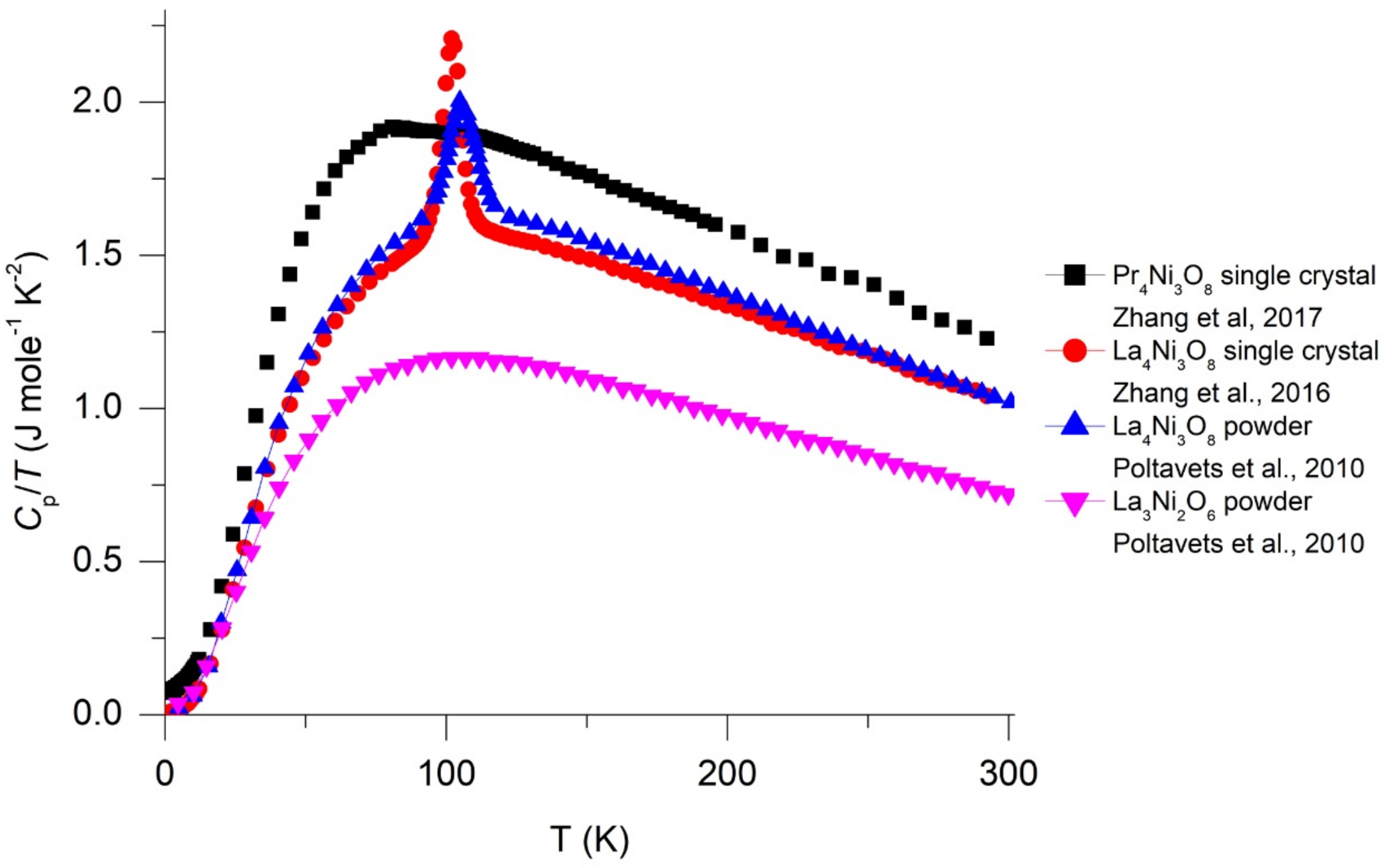

**Fig. 6** Heat capacity of $R_{n+1}Ni_nO_{2n+2}$ (R=La, Pr and Nd, n=2, 3, ∞). Zhang et al., 2017[13]; Zhang et al., 2016[15]; Poltavets et al., 2010[16]; Poltavets et al., 2010[16].

**Heat capacity**. Heat capacity data were reported for $La_4Ni_3O_8$ and $Pr_4Ni_3O_8$, and $La_3Ni_2O_6$ (see **Fig. 6**). An anomaly is clearly observed at ~105 K in both polycrystalline[16] and single crystals[15] of $La_4Ni_3O_8$, corresponding to the anomaly in resistivity and magnetic susceptibility. The entropy change across the transition was evaluated to be ~2 J mole$^{-1}$ K$^{-1}$ Ni$^{-1}$, which has been attributed to the condensation of the short range, fluctuating charge stripes[15]. In contrast, no anomaly was observed in the heat capacity of $Pr_4Ni_3O_8$ single crystals[13] or the bilayer $La_3Ni_2O_6$[16].

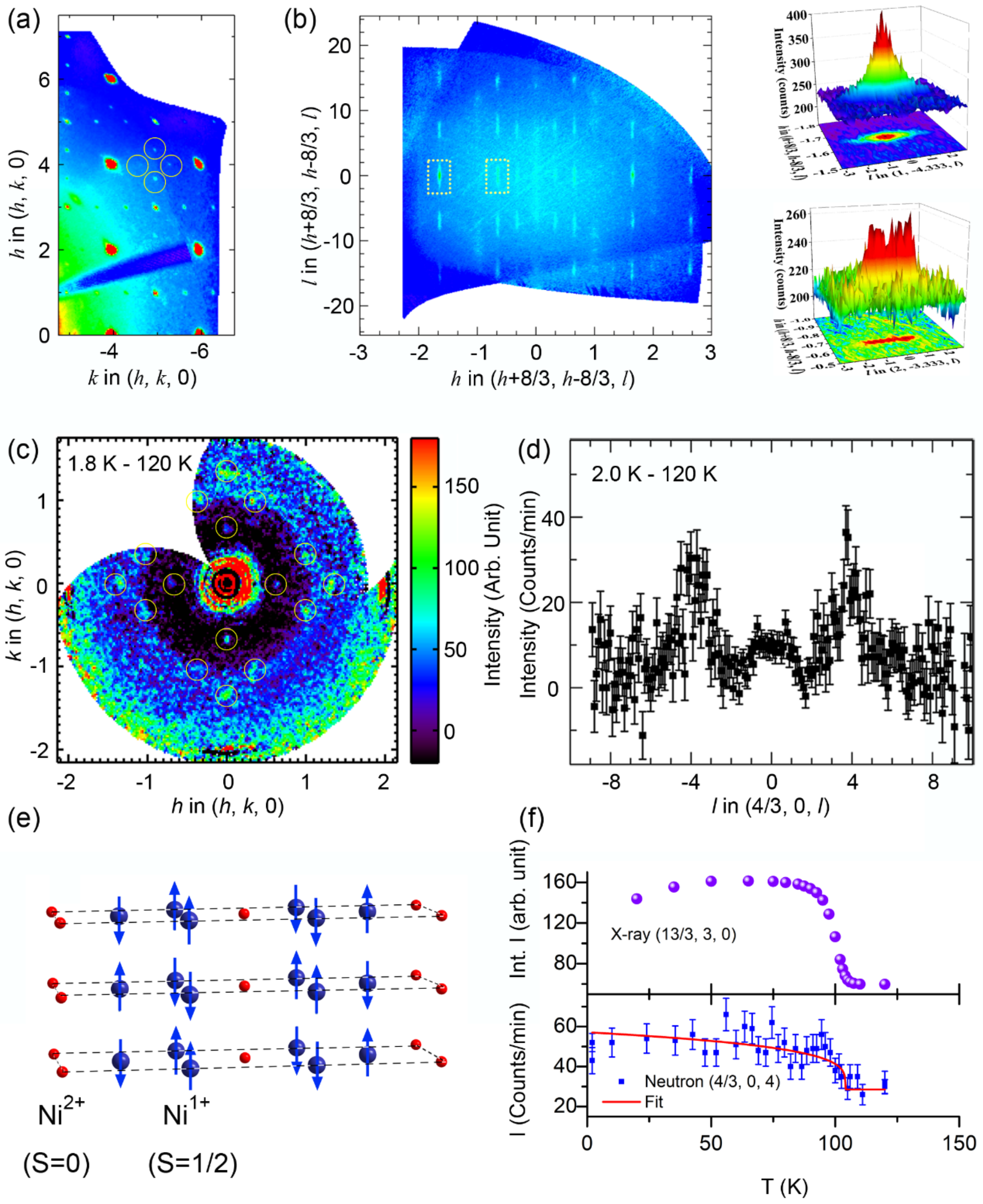

**Fig. 7** Charge and spin stripe order in the $La_4Ni_3O_8$[15, 33].

## 6. Cuprate-like physics in $R_4Ni_3O_8$ (R=La and Pr)

**Charge stripes in $La_4Ni_3O_8$.** Studies on polycrystalline powder samples of $La_4Ni_3O_8$ revealed a 105 K phase transition with a pronounced electronic, magnetic and lattice response[16]. The origin of the phase transition remained elusive[16, 38, 41] until Zhang et al. prepared single crystals in 2016[15]. Synchrotron X-ray single crystal diffraction revealed superlattice reflections locating at (1/3, 1/3, 0) on a basis of a tetragonal unit cell with a=b~3.97 Å when the samples were cooled below 105 K (see **Figs. 7a,b**). The intensity of such superlattice peaks is weaker in five orders of magnitude compared with the strongest Bragg peak, making it challenging to establish a quantitative model for the ground state. One explanation is charge stripe order considering the nominal $Ni^{1+}$: $Ni^{2+}$=2:1 and the similarity between $La_4Ni_3O_8$ and $La_{5/3}Sr_{1/3}NiO_4$[15, 116, 117]. Specifically, $La_4Ni_3O_8$ exhibits a semiconductor-to-insulating transition with a slope change in resistivity, a kink in magnetic susceptibility and an entropy change of 1.98 J $mol^{-1}$ $K^{-1}$ $Ni^{-1}$ across the phase transition. All these features are similar to those in the charge and spin stripe model system $La_{5/3}Sr_{1/3}NiO_4$[27, 118]. Additionally, the average oxidation state of Ni is 1.33+, separated by an integral charge from 2.33+ in $La_{5/3}Sr_{1/3}NiO_4$. Most importantly, the propagation vector for charge ordering obtained from single crystal diffraction is the same for $La_4Ni_3O_8$[15] and $La_{5/3}Sr_{1/3}NiO_4$[119].

The position and intensity distribution of the observed superlattice reflections provide information for the real-space arrangement of the charge stripes (see **Fig. 7e**)[15]. Although $La_4Ni_3O_8$ has the crystal structure of tetragonal (*I*4/*mmm*) whose unit cell is described by $a_t \times a_t \times c_t$, an orthorhombic setting with twice unit cell size $\sqrt{2}a_t \times \sqrt{2}a_t \times c_t$ is employed for easier comparison with single-layer nickelates[27, 120]. In this setting, the charge stripe superlattice propagation vector $\boldsymbol{q}$ is (2/3, 0, 1). Such a propagation vector indicates a threefold diagonal stripe pattern in the *ab* plane with charge stripes orientating at 45° to the Ni-O bonds, like single-layer nickelates as well as cobaltites and manganites[118]. The intensity distribution along *l*, e.g., maximum values at $l$=8$n$, signifies an in-phase stacking pattern of charge stripes within each trilayer, although stacking of charge is counterintuitive with a simple Coulomb repulsion model. The observation of distribution of intensity in the *ab* plane, i.e., ($h$±1/3, $k$±1, 0) are more intense than ($h$±2/3, $k$, 0), can be reproduced by a model with *ABAB* stacking of the stripes along *c*. Such a staggered pattern reduces the Coulomb repulsion among stripes. Additional support for the *ABAB* model is from different intensity distribution along *l* in the vicinity of the (1, -4.333, 0) and (2, -3.333, 0) reflections. The stripes are long range ordered in the *ab* plane but weakly correlated along *c* with an estimation of out-of-plane correlation length of 8 Å. Density functional theory (DFT)-based calculations verify a charge-ordered phase of $Ni^{1+}$ (S = 1/2)/$Ni^{2+}$(S = 0) stripes but find almost equal charge of the two types of Ni atoms[121]. Similar finding that different "charge states" of cations possess identical 3d orbital occupation (i.e., the charge) has been discussed in the past[122].

**Spin stripes in $La_4Ni_3O_8$**. The real-space ordering of charge into stripes found in $La_4Ni_3O_8$ provides an explanation for the semiconductor-to-insulator transition[15]; however, whether the ground state is magnetic remains an open question. **Fig. 7c** shows the difference between 1.8 K and 120 K data in the (*hk*0) scattering plane using unpolarized neutrons.[33] Weak peaks at (2/3, 0, 0), (4/3, 0, 0), (1, -1/3, 0), (1, 1/3, 0) and symmetrically equivalent positions are clearly observed. These superlattice peaks overlap with the charge peaks[15], so the key question is whether they have a magnetic component. Spin polarized measurements found that both nuclear and magnetic components contribute to (4/3, 0, 0). Additional evidence is from the *l* dependence of the neutron superlattice peaks as shown in **Fig. 7d**. Strong broad peaks locating at (4/3, 0, ±4) and one weak peak at (4/3, 0, 0) are clearly observed. The peak positions at $l$=±4 are apparently different from those observed ($l$=8n) in synchrotron X-ray single crystal diffraction[15], unambiguously demonstrating that (4/3, 0, ±4) are of magnetic origin. Further evidence is from spin polarized neutron diffraction, where a peak is clearly observed in the *h* scan through (4/3, 0, 4) in the spin flip channel at 1.5 K but not in the non-spin flip channel, nor in either channel at 120 K. For magnetic order, the propagation vector is measured from the antiferromagnetic wave vector (100) in the orthorhombic setting, which corresponds to the ($\pi$, $\pi$) point in the *I*4/*mmm* setting. Thus, $\boldsymbol{q}_{spin}$ is (1/3, 0, 0), which is one half of $\boldsymbol{q}_{charge}$=(2/3, 0, 0). The relationship between charge and spin order in $La_4Ni_3O_8$ satisfies $\boldsymbol{q}_{spin}=\frac{1}{2}\boldsymbol{q}_{charge}$, which signifies the existence of charge stripes as magnetic antiphase domain walls[118, 119].

The best spin stripe model consists of a spin axis along ***c***, antiferromagnetic coupling across charge stripes in the *ab* plane, antiferromagnetic interactions between layers within the trilayer, and uncorrelated between trilayers, as shown in **Fig. 7e**.[33] In this model, the spin stripes consist of $Ni^{+}$ with $S$=1/2 with charge stripes of $Ni^{2+}$ ($S$=0) as the domain walls, in good agreement with density functional theory calculations[121] and X-ray absorption spectroscopy[13]. Compared with quasi-2D oxides such as cuprates, single-layer nickelates, manganites and cobaltites, $La_4Ni_3O_8$ exhibits two unique features: (1) The spin direction is along ***c*** instead of in the *ab* plane, (2) the charge and spin stripe order occur at the same temperature (see **Fig. 7f**) while in other quasi-2D oxides spin order is the secondary order parameter[118, 123].

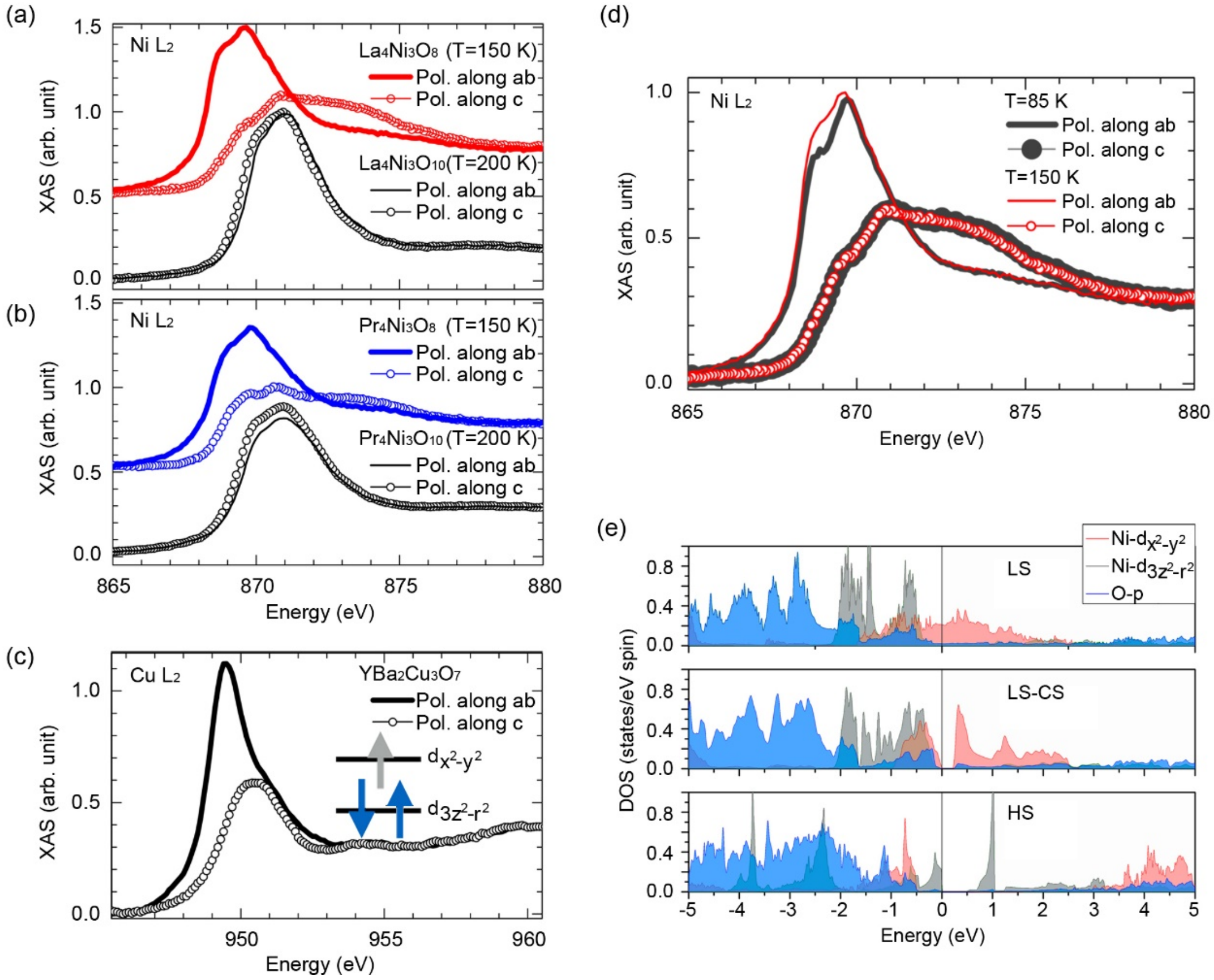

**Fig. 8** Large orbital polarization in $R_4Ni_3O_8$ (R=La, Pr)[33].

**Large orbital polarization in $La_4Ni_3O_8$.** The existence of quasi-2D square lattice[15], and charge[15] and spin stripes[33] in the ground state makes $La_4Ni_3O_8$ very close to the single-layer nickelates and cuprates[118]. One of the key questions is whether $La_4Ni_3O_8$ exists orbital polarization of the unoccupied $e_g$ states like that in the cuprates, rather than orbital degeneracy in the single-layer nickelates[117]. Structurally, the square planar trilayer nickelates such as $La_4Ni_3O_8$ are "natural" systems of sandwiched $LaNiO_3$-heterostructures with no apical oxygen around Ni. It is worth noting that $LaNiO_3$-heterostructures has been engineered for $x^2$-$y^2$ single band structure using strain, confinement, charge transfer and inversion symmetry breaking[19-21]. Indeed, a hole ratio ($h_{3z2\text{-}r2}/h_{x2\text{-}y2}$) of 0.55 has been obtained in the $LaTiO_3/LaNiO_3/LaAlO_3$ heterostructure, representing a ~50% reduction in the hole ratio compared to the bulk[20].

**Fig. 8a** presents the polarization dependent X-ray absorption spectra (XAS) at the Ni $L_2$ edge for $La_4Ni_3O_8$ with $La_4Ni_3O_{10}$ and $YBa_2Cu_3O_7$ as references (see **Fig. 8c**)[13]. A substantial difference in intensity near the leading part of the absorption edge is clearly seen between the XAS along *ab* and *c* in $La_4Ni_3O_8$, indicating a large orbital polarization of the unoccupied $e_g$ states. Such an observation is in sharp contrast to $La_4Ni_3O_{10}$ that shows orbital degeneracy, but resembles what has been found in $YBa_2Cu_3O_7$[124] which is considered to exhibit empty states of predominant $x^2$-$y^2$ character. The hole ratio ($h_{3z2\text{-}r2}/h_{x2\text{-}y2}$) in $La_4Ni_3O_8$ and $YBa_2Cu_3O_7$ was estimated by integrating the background-subtracted XAS data following a sum rule analysis[13]. Hole ratios of 0.5 and 0.35 were obtained for $La_4Ni_3O_8$ and $YBa_2Cu_3O_7$, respectively. These values significantly deviate from 1, indicating an orbital polarization favoring holes with $x^2$-$y^2$ character.

The polarization dependent XAS at the Ni $L_2$ edge for $La_4Ni_3O_8$ below the 105 K phase transition is shown in **Fig. 8d**.[13] Notably, there is only a slight change in the in-plane XAS and no change in the out-of-plane polarized XAS across the transition. The ground state thus shows large orbital polarization with dominant holes residing on the $x^2$-$y^2$ orbitals. This finding is inconsistent with the previous proposal of high-temperature low-spin to the low-temperature high spin state transition[38, 39], but agrees well with the low-spin state to low spin stripe state transition[15, 33]. The density of states for low-spin state, low-spin stripe state, and high-spin state are calculated in the framework of density functional theory (see **Fig. 8e**). In low-spin and low-spin stripe states, the unoccupied 3*d* states are calculated to be solely $x^2$-$y^2$ in character, in contrast to the $3z^2$-$r^2$ character in the high-spin state.[13]

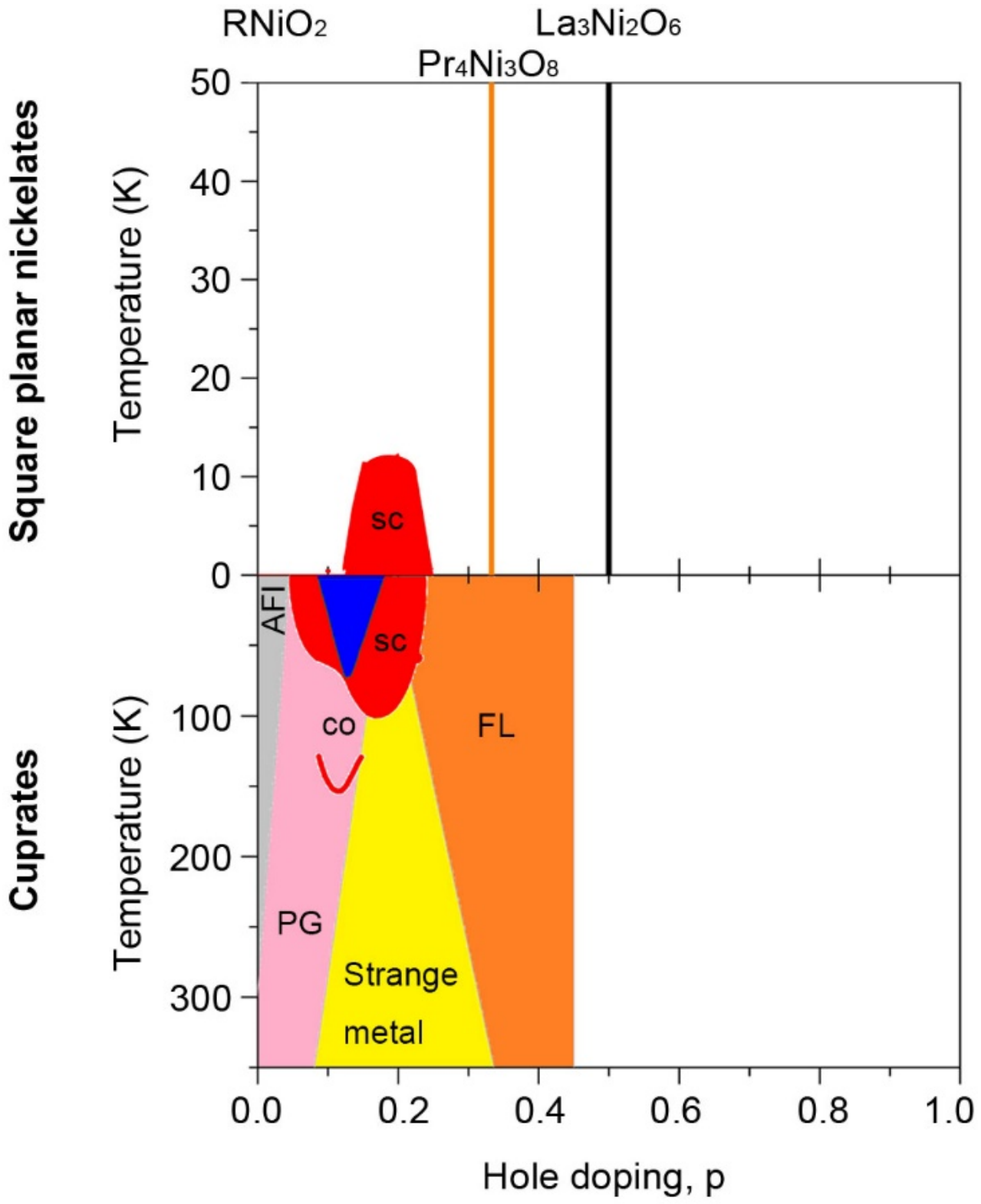

**Fig. 9** Schematic electronic phase diagram of quasi-2D square planar nickelates and cuprates as a function of the nominal 3d electron count[13, 69, 70].

**A close cuprate analogue: $Pr_4Ni_3O_8$**. As shown previously, the $La_4Ni_3O_8$ exhibits several key ingredients of cuprate superconductors, including a low-spin configuration with large orbital polarization and pronounced $x^2$-$y^2$ character in the unoccupied states near the Fermi level.[13] However, $La_4Ni_3O_8$ is not superconducting but semiconducting, and undergoes a semiconductor-to-insulator transition.[15] To become a cuprate analogue, 1/3 hole doped nickelates should be metallic, corresponding to $La_{1.67}Sr_{0.33}CuO_4$[125].

Hydrostatic pressure was utilized to tune the electronic structure of $La_4Ni_3O_8$[39, 41]. Cheng et al. successfully suppressed the insulator-metal transition at 6 GPa[41], consistent with DFT calculations by Prado et al.[39] However, a low-spin metallic phase does not emerge after the suppression. Instead, a structural reconstruction from fluorite La-$O_2$-La blocks at low pressure to rock-salt LaO-LaO blocks at high pressure was observed.[41]

Inspired by the DFT calculations by Pardo et al., Zhang et al. applied chemical pressure to tune the electronic structure of $R_4Ni_3O_8$[13]. According to Pardo et al., the insulator-metal transition occurs when the volume of unit cell of $La_4Ni_3O_8$ is reduced by 1%[39]. Considering the lanthanide contraction, the volume of $R_4Ni_3O_8$ shrinks from R=La to Nd. Single crystals of $Pr_4Ni_3O_8$ were successfully prepared by Zhang et al.[13, 32] At room temperature, the volume of $Pr_4Ni_3O_8$ is 96% of $La_4Ni_3O_8$, falling into the metallic region of $La_4Ni_3O_8$ under high pressure calculated by Pardo et al.[39] Indeed, $Pr_4Ni_3O_8$ is metallic as the resistivity decreases when the sample is cooled (see **Fig. 4**). In addition, no phase transition was observed down to 2 K, indicating that static charge and spin stripe phase is suppressed (see **Figs. 4-6**).[13] What about the spin state of Ni and orbital polarization in $Pr_4Ni_3O_8$? The polarization dependent XAS at the Ni $L_2$ edge at 150 K is shown in **Fig. 8b**. Same to $La_4Ni_3O_8$, $Pr_4Ni_3O_8$ shows a low-spin configuration with significant orbital polarization and dominant $x^2$-$y^2$ character in the unoccupied states above the Fermi energy[13].

**Fig. 9** presents a schematic electronic phase diagram of quasi-2D square planar nickelates and cuprates as a function of the nominal 3d electron count. The bottom shows the generic phase diagram of cuprates which consists of antiferromagnetic insulator (AFI), superconducting phase (SC), strange metal, pseudogap (PG), Fermi liquid (FL), and charge/spin ordering (CO).[8] Compared with cuprates, the quasi-2D square planar nickelates is less explored.[13] Only several line compounds have been studied, including $RNiO_2$ (R=La, Nd; $d^9$)[28, 34, 35, 80, 81], $La_3Ni_2O_6$ ($d^{8.5}$)[17, 76], and $R_4Ni_3O_8$ (R=La, Pr, Nd; $d^{8.67}$)[13, 15, 16, 33, 37, 39, 79, 84, 94, 97, 99]. Recently, a superconducting dome in the range of $d^{8.88}$-$d^{8.72}$ has been reported in thin films of $R_{1-x}Sr_xNiO_2$ (R=Pr, Nd),[51, 69-72] which is very close to the superconducting dome of the curpates.[8] The trilayer nickelate $Pr_4Ni_3O_8$, with 1/3 self hole doping and lying in the d-count regime corresponding to the overdoped Fermi liquid of the cuprates, is a potential material for exploring superconductivity if electron doping can be achieved.[13]

Theoretical calculations predict superconductivity in the electron-doped trilayer nickelates. Recently, theoretical calculations by Botana et al. show that Ce-doped $Pr_4Ni_3O_8$ ($Pr_3CeNi_3O_8$) has the same electronic structure as the antiferromagnetic insulating phase of parent cuprates[14]. Combining first-principles and *t-J* model calculations, Nica et al. found that $R_4Ni_3O_8$ resemble cuprates more than $RNiO_2$ materials in that only Ni $d_{x^2-y^2}$ bands cross the Fermi level and the superexchange interactions are significantly enhanced[126]. They predicted a maximum superconducting transition temperature of ~ 90 K in $Pr_3CeNi_3O_8$[126], much larger than the 15 K observed in $Nd_{0.8}Sr_{0.2}NiO_2$ thin film[51].

## 7. Summary and outlook

In summary, we reviewed the synthesis, structure, crystal growth and physical properties of the quasi-2D square planar nickelates, a fast-moving field accelerated by the discovery of superconductivity in thin films. We highlighted the single crystal growth of the trilayer nickelates $R_4Ni_3O_{10}$ (R=La and Pr) using the high-pressure floating zone technique, and the following topotactic reduction to $R_4Ni_3O_8$ (R=La and Pr) single crystals, which are crucial for revealing their cuprate-like physics including charge and spin stripes, large orbital polarization, low spin state, and significant *p-d* hybridization.

For the future perspective of layered nickelates, the following aspects should be addressed:

First, high-quality single crystals of $R_{1-x}Sr_xNiO_2$ (R=Pr and Nd) are needed to address the fundamental question whether superconductivity is intrinsic to layered nickelates. In the heterostructures, there exist stress/strain and interface between $R_{1-x}Sr_xNiO_2$ (R=Pr and Nd) thin film and the substrate such as $SrTiO_3$. While for bulk polycrystalline samples of $R_{1-x}Sr_xNiO_2$ (R=Pr and Nd), they were found to be insulating rather than superconducting. It is worth noting that the powder samples contain a small amount of impurities. Since powders and thin films of $R_{1-x}Sr_xNiO_2$ (R=Pr and Nd) have been successfully prepared via the topotactic reduction, it is natural to follow the procedure to prepare their single crystals via perovskite crystals. The successful transformation from $R_4Ni_3O_{10}$ (R=La and Pr) bulk single crystals to $R_4Ni_3O_8$ (R=La and Pr) single crystals with dimensions of 1~2 $mm^2$×0.5 mm has proved the feasibility of topotactic reduction on single crystals[13]. $RNiO_3$ (R=Pr-Lu) crystallize in the orthorhombic space group *Pbnm* above their metal-insulator transition temperatures, in which *a* and *b* axes are *different*. The high-temperature structure of perovskites is possibly $Pm\overline{3}m$, which is supergroup of *Pbnm*. Twining is highly possible during the structural transition from cubic $Pm\overline{3}m$ to orthorhombic *Pbnm*. Therefore, selection of an untwined crystal of $R_{1-x}Sr_xNiO_3$ (R=Pr and Nd) is important in order to obtain one single-domain crystal of $R_{1-x}Sr_xNiO_2$ (R=Pr and Nd) via topotactic reduction. Indeed, untwined crystal of $NdNiO_3$ has been reported by Alonso et al.[127] In addition, detwinning techniques such as application of uniaxial stress/strain at high temperature may be utilized; these techniques have been successfully used in obtaining single-domain crystal of iron-based superconductors[128, 129] and cuprates[130].

Second, single crystal growth of $R_{1-x}Sr_xNiO_3$ (R=Pr and Nd) and $R_{1-x}Sr_xNiO_2$ (R=Pr and Nd) is challenging. Two methods are promising: (i) high pressure flux method, and (ii) high pressure floating zone techniques. For (i), it is difficult to grow large single crystals. For (ii), oxygen pressure of more than 30 and 220 bar are needed to stabilize $LaNiO_3$ and $PrNiO_3$, respectively. Hole doping in $PrNiO_3$, for example substituting Pr with Sr to form $Pr_{1-x}Sr_xNiO_3$, increase the oxidization state of Ni to 3+*x*. As demonstrated by $La_{n+1}Ni_nO_{3n+1}$, the high oxidization state requires higher oxygen pressure to stabilize the target phase. Thus, it is expected to be challenge to stabilize $Pr_{1-x}Sr_xNiO_3$. It is likely that more than 300 bar oxygen pressure is needed to stabilize

$NdNiO_3$ and $Nd_{1-x}Sr_xNiO_3$.

Third, the magnetic ground state of $RNiO_2$ (R=La, Pr) remains an open question. Single crystals of $RNiO_3$ (R=La, Pr) has been reported, so it is possible to prepare single crystals of $RNiO_2$ (R=La, Pr) and perform neutron scattering experiments to determine their ground state is magnetic or not. However, the preparation of $LaNiO_2$ single crystals may be challenging because there is no preferential layering direction in $LaNiO_3$ (very small distortion from the cubic phase), the likelihood is a multidomain specimen in which three possible layering directions are completely homogenized.

Finally, tuning the electronic structure of quasi-2D square planar nickelates by varying Ni-O layers. (1) electron doping in the cuprate analogue $Pr_4Ni_3O_8$ might host high-$T_c$ superconductivity; (2) the 3d electron count of $R_{n+1}Ni_nO_{2n+2}$ (n=4-8) falls in the range of the superconducting dome found in thin films of $R_{1-x}Sr_xNiO_2$ (R=Pr and Nd). The preparation of their parent single crystals of $R_{n+1}Ni_nO_{3n+1}$ (n=4-8) is probably less challenging compared with $R_{1-x}Sr_xNiO_3$ considering the Ni valence of $(3-\frac{1}{n})$ and the absence of disorder problem associated with Sr doping.

## Conflicts of interest

The authors declare no competing financial interest.

## Acknowledgements

We gratefully acknowledge financial support from the National Key Research and Development Program of China (Grant No. 2016YFB1102201, 2018YFB0406502), the National Natural Science Foundation of China (Grant No. 51932004, 61975098, 12074219), the 111 Project 2.0 (Grant No: BP2018013), the Qilu Young Scholars Program of Shandong University, and the Taishan Scholars Program of Shandong Province.

## References

1. N. Plakida, *High-Temperature Cuprate Superconductors: Experiment, Theory, and Application*, Springer, Heidelberg, 2010.
2. D. v. Delft and P. Kes, *Phys. Today*, 2010, **63**, 38.
3. A. Schilling, M. Cantoni, J. D. Guo and H. R. Ott, *Nature*, 1993, **363**, 56-58.
4. E. Snider, N. Dasenbrock-Gammon, R. McBride, M. Debessai, H. Vindana, K. Vencatasamy, K. V. Lawler, A. Salamat and R. P. Dias, *Nature*, 2020, **586**, 373-377.
5. M. Somayazulu, M. Ahart, A. K. Mishra, Z. M. Geballe, M. Baldini, Y. Meng, V. V. Struzhkin and R. J. Hemley, *Phys. Rev. Lett.*, 2019, **122**, 027001.
6. A. P. Drozdov, P. P. Kong, V. S. Minkov, S. P. Besedin, M. A. Kuzovnikov, S. Mozaffari, L. Balicas, F. F. Balakirev, D. E. Graf, V. B. Prakapenka, E. Greenberg, D. A. Knyazev, M. Tkacz and M. I. Eremets, *Nature*, 2019, **569**, 528-531.
7. A. P. Drozdov, M. I. Eremets, I. A. Troyan, V. Ksenofontov and S. I. Shylin, *Nature*, 2015, **525**, 73.
8. B. Keimer, S. A. Kivelson, M. R. Norman, S. Uchida and J. Zaanen, *Nature*, 2015, **518**, 179-186.
9. Q. Si, R. Yu and E. Abrahams, *Nat. Rev. Mater.*, 2016, **1**, 16017.
10. Y. K. Kim, O. Krupin, J. D. Denlinger, A. Bostwick, E. Rotenberg, Q. Zhao, J. F. Mitchell, J. W. Allen and B. J. Kim, *Science*, 2014, **345**, 187-190.
11. J. Bertinshaw, Y. K. Kim, G. Khaliullin and B. J. Kim, *Annu. Rev. Condens. Matter Phys.*, 2019, **10**, 315-336.
12. J. Gawraczyński, D. Kurzydłowski, R. A. Ewings, S. Bandaru, W. Gadomski, Z. Mazej, G. Ruani, I. Bergenti, T. Jaroń, A. Ozarowski, S. Hill, P. J. Leszczyński, K. Tokár, M. Derzsi, P. Barone, K. Wohlfeld, J. Lorenzana and W. Grochala, *Proc. Natl. Acad. Sci. U. S. A.*, 2019, **116**, 1495-1500.
13. J. Zhang, A. S. Botana, J. W. Freeland, D. Phelan, H. Zheng, V. Pardo, M. R. Norman and J. F. Mitchell, *Nat. Phys.*, 2017, **13**, 864-869.

14. A. S. Botana, V. Pardo and M. R. Norman, *Phys. Rev. Mater.*, 2017, **1**, 021801.
15. J. Zhang, Y. S. Chen, D. Phelan, H. Zheng, M. R. Norman and J. F. Mitchell, *Proc. Natl. Acad. Sci. U.S.A.*, 2016, **113**, 8945-8950.
16. V. V. Poltavets, K. A. Lokshin, A. H. Nevidomskyy, M. Croft, T. A. Tyson, J. Hadermann, G. Van Tendeloo, T. Egami, G. Kotliar, N. ApRoberts-Warren, A. P. Dioguardi, N. J. Curro and M. Greenblatt, *Phys. Rev. Lett.*, 2010, **104**, 206403.
17. V. V. Poltavets, M. Greenblatt, G. H. Fecher and C. Felser, *Phys. Rev. Lett.*, 2009, **102**, 046405.
18. V. I. Anisimov, D. Bukhvalov and T. M. Rice, *Phys. Rev. B*, 1999, **59**, 7901-7906.
19. A. S. Disa, F. J. Walker, S. Ismail-Beigi and C. H. Ahn, *APL Mater.*, 2015, **3**, 062303.
20. A. S. Disa, D. P. Kumah, A. Malashevich, H. Chen, D. A. Arena, E. D. Specht, S. Ismail-Beigi, F. J. Walker and C. H. Ahn, *Phys. Rev. Lett.*, 2015, **114**, 026801.
21. P. Hansmann, X. Yang, A. Toschi, G. Khaliullin, O. K. Andersen and K. Held, *Phys. Rev. Lett.*, 2009, **103**, 016401.
22. J. Chaloupka and G. Khaliullin, *Phys. Rev. Lett.*, 2008, **100**, 016404.
23. L. Zhao, D. H. Torchinsky, H. Chu, V. Ivanov, R. Lifshitz, R. Flint, T. Qi, G. Cao and D. Hsieh, *Nat Phys*, 2016, **12**, 32-36.
24. Y. K. Kim, N. H. Sung, J. D. Denlinger and B. J. Kim, *Nat. Phys.*, 2016, **12**, 37-41.
25. Y. J. Yan, M. Q. Ren, H. C. Xu, B. P. Xie, R. Tao, H. Y. Choi, N. Lee, Y. J. Choi, T. Zhang and D. L. Feng, *Phys. Rev. X*, 2015, **5**, 041018.
26. X. Yang and H. Su, *Sci. Rep.*, 2014, **4**, 5420.
27. J. M. Tranquada, *AIP Conf. Proc.*, 2013, **1550**, 114-187.
28. M. Crespin, P. Levitz and L. Gatineau, *J. Chem. Soc., Faraday Trans. 2*, 1983, **79**, 1181-1194.
29. J. A. Alonso, A. Muñoz, A. Largeteau and G. Demazeau, *J. Phys.: Condens. Matter*, 2004, **16**, S1277.
30. T. Saito, M. Azuma, E. Nishibori, M. Takata, M. Sakata, N. Nakayama, T. Arima, T. Kimura, C. Urano and M. Takano, *Physica B*, 2003, **329–333, Part 2**, 866-867.
31. J. Zhang, H. Zheng, Y. Ren and J. F. Mitchell, *Cryst. Growth Des.*, 2017, **17**, 2730-2735.
32. J. Zhang, H. Zheng, Y.-S. Chen, Y. Ren, M. Yonemura, A. Huq and J. F. Mitchell, *Phys. Rev. Mater.*, 2020, **4**, 083402.
33. J. Zhang, D. M. Pajerowski, A. S. Botana, H. Zheng, L. Harriger, J. Rodriguez-Rivera, J. P. C. Ruff, N. J. Schreiber, B. Wang, Y.-S. Chen, W. C. Chen, M. R. Norman, S. Rosenkranz, J. F. Mitchell and D. Phelan, *Phys. Rev. Lett.*, 2019, **122**, 247201.
34. P. Levitz, M. Crespin and L. Gatineau, *J. Chem. Soc., Faraday Trans. 2*, 1983, **79**, 1195-1203.
35. M. A. Hayward, M. A. Green, M. J. Rosseinsky and J. Sloan, *J. Am. Chem. Soc.*, 1999, **121**, 8843-8854.
36. K. W. Lee and W. E. Pickett, *Phys. Rev. B*, 2004, **70**, 165109.
37. P. Lacorre, *J. Solid State Chem.*, 1992, **97**, 495-500.
38. V. Pardo and W. E. Pickett, *Phys. Rev. Lett.*, 2010, **105**, 266402.
39. V. Pardo and W. E. Pickett, *Phys. Rev. B*, 2012, **85**, 045111.
40. N. ApRoberts-Warren, A. P. Dioguardi, V. V. Poltavets, M. Greenblatt, P. Klavins and N. J. Curro, *Phys. Rev. B*, 2011, **83**, 014402.
41. J. G. Cheng, J. S. Zhou, J. B. Goodenough, H. D. Zhou, K. Matsubayashi, Y. Uwatoko, P. P. Kong, C. Q. Jin, W. G. Yang and G. Y. Shen, *Phys. Rev. Lett.*, 2012, **108**, 236403.
42. H. Zheng, B.-X. Wang, D. Phelan, J. Zhang, Y. Ren, M. J. Krogstad, S. Rosenkranz, R. Osborn and J. F. Mitchell, *Crystals*, 2020, **10**, 557.
43. J. Zhang, D. Phelan, A. S. Botana, Y.-S. Chen, H. Zheng, M. Krogstad, S. G. Wang, Y. Qiu, J. A. Rodriguez-Rivera, R. Osborn, S. Rosenkranz, M. R. Norman and J. F. Mitchell, *Nat. Commun.*, 2020, **11**, 6003.
44. H. Zheng, J. Zhang, B. Wang, D. Phelan, M. J. Krogstad, Y. Ren, W. A. Phelan, O. Chmaissem, B. Poudel and J. F. Mitchell, *Crystals*, 2019, **9**, 324.
45. V. L. Karner, A. Chatzichristos, D. L. Cortie, M. H. Dehn, O. Foyevtsov, K. Foyevtsova, D. Fujimoto, R. F. Kiefl, C. D. P. Levy, R. Li, R. M. L. McFadden, G. D. Morris, M. R. Pearson, M. Stachura, J. O. Ticknor, G. Cristiani, G. Logvenov, F. Wrobel, B. Keimer, J. Zhang, J. F. Mitchell and W. A. MacFarlane, *Phys. Rev. B*, 2019, **100**, 165109.
46. B.-X. Wang, S. Rosenkranz, X. Rui, J. Zhang, F. Ye, H. Zheng, R. F. Klie, J. F. Mitchell and D. Phelan, *Phys. Rev. Mater.*, 2018, **2**, 064404.

47. H. Li, X. Zhou, T. Nummy, J. Zhang, V. Pardo, W. E. Pickett, J. F. Mitchell and D. S. Dessau, *Nat. Commun.*, 2017, **8**, 704.
48. H. Guo, Z. W. Li, L. Zhao, Z. Hu, C. F. Chang, C. Y. Kuo, W. Schmidt, A. Piovano, T. W. Pi, O. Sobolev, D. I. Khomskii, L. H. Tjeng and A. C. Komarek, *Nat. Commun.*, 2018, **9**, 43.
49. S. Huangfu, G. D. Jakub, X. Zhang, O. Blacque, P. Puphal, E. Pomjakushina, F. O. v. Rohr and A. Schilling, *Phys. Rev. B*, 2020, **101**, 104104.
50. K. Dey, W. Hergett, P. Telang, M. M. Abdel-Hafiez and R. Klingeler, *J. Cryst. Growth*, 2019, **524**, 125157.
51. D. Li, K. Lee, B. Y. Wang, M. Osada, S. Crossley, H. R. Lee, Y. Cui, Y. Hikita and H. Y. Hwang, *Nature*, 2019, **572**, 624-627.
52. Y.-H. Zhang and A. Vishwanath, *Phys. Rev. Research*, 2020, **2**, 023112.
53. H. Zhang, L. Jin, S. Wang, B. Xi, X. Shi, F. Ye and J.-W. Mei, *Phys. Rev. Research*, 2020, **2**, 013214.
54. X. Wu, D. D. Sante, T. Schwemmer, W. Hanke, H. Y. Hwang, S. Raghu and R. Thomale, *Phys. Rev. B*, 2020, **101**, 060504(R).
55. P. Werner and S. Hoshino, *Phys. Rev. B*, 2020, **101**, 041104(R).
56. H. Sakakibara, H. Usui, K. Suzuki, T. Kotani, H. Aoki and K. Kuroki, *Phys. Rev. Lett.*, 2020, **125**, 077003.
57. S. Ryee, H. Yoon, T. J. Kim, M. Y. Jeong and M. J. Han, *Phys. Rev. B*, 2020, **101**, 064513.
58. V. Pardo and A. S. Botana, 2020, Preprint at https://arxiv.org/abs/2012.02711.
59. I. Leonov, S. L. Skornyakov and S. Y. Savrasov, *Phys. Rev. B*, 2020, **101**, 241108.
60. F. Lechermann, *Phys. Rev. B*, 2020, **101**, 081110(R).
61. M. Kitatani, L. Si, O. Janson, R. Arita, Z. Zhong and K. Held, *npj Quantum Mater.*, 2020, **5**, 59.
62. J. Karp, A. S. Botana, M. R. Norman, H. Park, M. Zingl and A. Millis, *Phys. Rev. X*, 2020, **10**, 021061.
63. M. Jiang, M. Berciu and G. A. Sawatzky, *Phys. Rev. Lett.*, 2020, **124**, 207004.
64. M. Hepting, D. Li, C. J. Jia, H. Lu, E. Paris, Y. Tseng, X. Feng, M. Osada, E. Been, Y. Hikita, Y.-D. Chuang, Z. Hussain, K. J. Zhou, A. Nag, M. Garcia-Fernandez, M. Rossi, H. Y. Huang, D. J. Huang, Z. X. Shen, T. Schmitt, H. Y. Hwang, B. Moritz, J. Zaanen, T. P. Devereaux and W. S. Lee, *Nat. Mater.*, 2020, **19**, 381-385.
65. Y. Gu, S. Zhu, X. Wang, J. Hu and H. Chen, *Comms. Phys.*, 2020, **3**, 84.
66. J. Gao, S. Peng, Z. Wang, C. Fang and H. Weng, *Natl. Sci. Rev.*, 2020, nwaa218.
67. A. S. Botana and M. R. Norman, *Phys. Rev. X*, 2020, **10**, 011024.
68. J. E. Hirsch and F. Marsiglio, *Physica C*, 2019, **566**, 1353534.
69. D. Li, B. Y. Wang, K. Lee, S. P. Harvey, M. Osada, B. H. Goodge, L. F. Kourkoutis and H. Y. Hwang, *Phys. Rev. Lett.*, 2020, **125**, 027001.
70. S. Zeng, C. S. Tang, X. Yin, C. Li, M. Li, Z. Huang, J. Hu, W. Liu, G. J. Omar, H. Jani, Z. S. Lim, K. Han, D. Wan, P. Yang, S. J. Pennycook, A. T. S. Wee and A. Ariando, *Phys. Rev. Lett.*, 2020, **125**, 147003.
71. M. Osada, B. Y. Wang, K. Lee, D. Li and H. Y. Hwang, *Phys. Rev. Mater.*, 2020, **4**, 121801.
72. M. Osada, B. Y. Wang, B. H. Goodge, K. Lee, H. Yoon, K. Sakuma, D. Li, M. Miura, L. F. Kourkoutis and H. Y. Hwang, *Nano Lett.*, 2020, **20**, 5735-5740.
73. Q. Li, C. He, J. Si, X. Zhu, Y. Zhang and H.-H. Wen, *Comms. Mater.*, 2020, **1**, 16.
74. B.-X. Wang, H. Zheng, E. Krivyakina, O. Chmaissem, P. P. Lopes, J. W. Lynn, L. C. Gallington, Y. Ren, S. Rosenkranz, J. F. Mitchell and D. Phelan, *Phys. Rev. Mater.*, 2020, **4**, 084409.
75. G. Behr, W. Löser, N. Wizent, P. Ribeiro, M. O. Apostu and D. Souptel, *J. Mater. Sci.*, 2010, **45**, 2223-2227.
76. V. V. Poltavets, K. A. Lokshin, S. Dikmen, M. Croft, T. Egami and M. Greenblatt, *J. Am. Chem. Soc.*, 2006, **128**, 9050-9051.
77. M. Azuma, T. Saito, S. Ishiwata, I. Yamada, Y. Kohsaka, H. Takagi and M. Takano, *Physica C*, 2003, **392–396, Part 1**, 22-28.
78. J. G. Bednorz and K. A. Müller, *Z. Phys. B: Condens. Matter*, 1986, **64**, 189-193.
79. V. V. Poltavets, K. A. Lokshin, M. Croft, T. K. Mandal, T. Egami and M. Greenblatt, *Inorg. Chem.*, 2007, **46**, 10887-10891.
80. M. Crespin, O. Isnard, F. Dubois, J. Choisnet and P. Odier, *J. Solid State Chem.*, 2005, **178**, 1326-1334.

81. M. A. Hayward and M. J. Rosseinsky, *Solid State Sci.*, 2003, **5**, 839-850.
82. M. J. Martínez-Lope, M. T. Casais and J. A. Alonso, *J. Alloys Compd.*, 1998, **275-277**, 109-112.
83. C. K. Blakely, S. R. Bruno and V. V. Poltavets, *Inorg. Chem.*, 2011, **50**, 6696-6700.
84. Q. Li, C. He, X. Zhu, J. Si, X. Fan and H.-H. Wen, *Sci. China-Phys. Mech. Astron.*, 2020, **64**, 227411.
85. G. Demazeau, A. Marbeuf, M. Pouchard and P. Hagenmuller, *J. Solid State Chem.*, 1971, **3**, 582-589.
86. A. Wold, B. Post and E. Banks, *J. Am. Chem. Soc.*, 1957, **79**, 4911-4913.
87. A. Wold, R. J. Arnott and J. B. Goodenough, *J. Appl. Phys.*, 1958, **29**, 387-389.
88. X. Q. Xu, J. L. Peng, Z. Y. Li, H. L. Ju and R. L. Greene, *Phys. Rev. B*, 1993, **48**, 1112-1118.
89. P. Lacorre, J. B. Torrance, J. Pannetier, A. I. Nazzal, P. W. Wang and T. C. Huang, *J. Solid State Chem.*, 1991, **91**, 225-237.
90. A. K. Norman and M. A. Morris, *J. Mater. Process. Technol.*, 1999, **92–93**, 91-96.
91. J. K. Vassiliou, M. Hornbostel, R. Ziebarth and F. J. Disalvo, *J. Solid State Chem.*, 1989, **81**, 208-216.
92. C. He, X. Ming, Q. Li, X. Zhu, J. Si and H.-H. Wen, 2020, Preprint at https://arxiv.org/abs/2010.11777.
93. M. D. Carvalho, F. M. A. Costa, I. D. S. Pereira, A. Wattiaux, J. M. Bassat, J. C. Grenier and M. Pouchard, *J. Mater. Chem.*, 1997, **7**, 2107-2111.
94. R. Retoux, J. Rodriguez-Carvajal and P. Lacorre, *J. Solid State Chem.*, 1998, **140**, 307-315.
95. N. P. Armitage, P. Fournier and R. L. Greene, *Rev. Mod. Phys.*, 2010, **82**, 2421-2487.
96. C. D. Ling, D. N. Argyriou, G. Wu and J. J. Neumeier, *J. Solid State Chem.*, 2000, **152**, 517-525.
97. S. Huangfu, Z. Guguchia, D. Cheptiakov, X. Zhang, H. Luetkens, D. J. Gawryluk, T. Shang, F. O. von Rohr and A. Schilling, *Phys. Rev. B*, 2020, **102**, 054423.
98. T. Duffar, *Crystal Growth Processes Based on Capillarity: Czochralski, Floating Zone, Shaping and Crucible Techniques*, WILEY, United Kingdom, 2010.
99. M. H. Upton, J. Zhang, H. Zheng, A. Said and J. F. Mitchell, *J. Phys.: Condens. Matter*, 2020, **32**, 425503.
100. J. Zhang, H. Zheng, C. D. Malliakas, J. M. Allred, Y. Ren, Q. A. Li, T. H. Han and J. F. Mitchell, *Chem. Mater.*, 2014, **26**, 7172-7182.
101. G. Ryu, H. Guo, L. Zhao, M. T. Fernández-Díaz, Y. Drees, Z. W. Li, Z. Hu and A. C. Komarek, *Phys. Status Solidi RRL*, 2019, **13**, 1800537.
102. P. Puphal, S. Allenspach, C. Rüegg and E. Pomjakushina, *Crystals*, 2019, **9**, 273.
103. H. Guo, Z. Hu, T.-W. Pi, L. Tjeng and A. Komarek, *Crystals*, 2016, **6**, 98.
104. H. B. Cao, Z. Y. Zhao, M. Lee, E. S. Choi, M. A. McGuire, B. C. Sales, H. D. Zhou, J. Q. Yan and D. G. Mandrus, *APL Mater.*, 2015, **3**, 062512.
105. W. A. Phelan, J. Zahn, Z. Kennedy and T. M. McQueen, *J. Solid State Chem.*, 2018, **270**, 705-709.
106. J. L. Schmehr, M. Aling, E. Zoghlin and S. D. Wilson, *Rev. Sci. Instrum.*, 2019, **90**, 043906.
107. A. Ikeda, T. Manabe and M. Naito, *Physica C*, 2014, **506**, 83-86.
108. Y. Sakurai, N. Chiba, Y. Kimishima and M. Uehara, *Physica C*, 2013, **487**, 27-30.
109. H. B. Radousky, *J. Mater. Res.*, 1992, **7**, 1917-1955.
110. I. Ai, K. Yoshiharu, I. Hiroshi, N. Michio and Y. Hideki, *Appl. Phys. Express*, 2016, **9**, 061101.
111. M. Kawai, K. Matsumoto, N. Ichikawa, M. Mizumaki, O. Sakata, N. Kawamura, S. Kimura and Y. Shimakawa, *Cryst. Growth Des.*, 2010, **10**, 2044-2046.
112. Q. A. Li, K. E. Gray and J. F. Mitchell, *Phys. Rev. B*, 1999, **59**, 9357-9361.
113. V. Hardy, A. Maignan, C. Martin, F. Warmont and J. Provost, *Phys. Rev. B*, 1997, **56**, 130-133.
114. O. O. Bernal, D. E. MacLaughlin, G. D. Morris, P. C. Ho, L. Shu, C. Tan, J. Zhang, Z. Ding, K. Huang and V. V. Poltavets, *Phys. Rev. B*, 2019, **100**, 125142.
115. F. Serrano-Sánchez, F. Fauth, J. L. Martínez and J. A. Alonso, *Inorg. Chem.*, 2019, **58**, 11828.
116. Y. Ikeda, S. Suzuki, T. Nakabayashi, H. Yoshizawa, T. Yokoo and S. Itoh, *J. Phys. Soc. Jpn.*, 2015, **84**, 023706.
117. R. Klingeler, B. Büchner, S. W. Cheong and M. Hücker, *Phys. Rev. B*, 2005, **72**, 104424.
118. H. Ulbrich and M. Braden, *Physica C*, 2012, **481**, 31-45.
119. S. H. Lee and S. W. Cheong, *Phys. Rev. Lett.*, 1997, **79**, 2514-2517.
120. J. M. Tranquada, B. J. Sternlieb, J. D. Axe, Y. Nakamura and S. Uchida, *Nature*, 1995, **375**, 561-563.

121. A. S. Botana, V. Pardo, W. E. Pickett and M. R. Norman, *Phys. Rev. B*, 2016, **94**, 081105(R).
122. Y. Quan, V. Pardo and W. E. Pickett, *Phys. Rev. Lett.*, 2012, **109**, 216401.
123. M. Hücker, M. v. Zimmermann, G. D. Gu, Z. J. Xu, J. S. Wen, G. Xu, H. J. Kang, A. Zheludev and J. M. Tranquada, *Phys. Rev. B*, 2011, **83**, 104506.
124. D. G. Hawthorn, K. M. Shen, J. Geck, D. C. Peets, H. Wadati, J. Okamoto, S. W. Huang, D. J. Huang, H. J. Lin, J. D. Denlinger, R. Liang, D. A. Bonn, W. N. Hardy and G. A. Sawatzky, *Phys. Rev. B*, 2011, **84**, 075125.
125. R. A. Cooper, Y. Wang, B. Vignolle, O. J. Lipscombe, S. M. Hayden, Y. Tanabe, T. Adachi, Y. Koike, M. Nohara, H. Takagi, C. Proust and N. E. Hussey, *Science*, 2009, **323**, 603-607.
126. E. M. Nica, J. Krishna, R. Yu, Q. Si, A. S. Botana and O. Erten, *Phys. Rev. B*, 2020, **102**, 020504(R).
127. J. A. Alonso, G. Demazeau, A. Largeteau, D. Kurowski, R.-D. Hoffmann and R. Pottgen, *Z. Naturforsch. B.*, 2006, **61**, 346-349.
128. P. Dai, *Rev. Mod. Phys.*, 2015, **87**, 855-896.
129. J.-H. Chu, J. G. Analytis, K. De Greve, P. L. McMahon, Z. Islam, Y. Yamamoto and I. R. Fisher, *Science*, 2010, **329**, 824-826.
130. T. Suzuki, S. Awaji, H. Oguro and K. Watanabe, *IEEE Transactions on Applied Superconductivity*, 2015, **25**, 1-4.